\documentclass[]{interact}

\usepackage{epstopdf}
\usepackage[caption=false]{subfig}

\usepackage[natbibapa,nodoi]{apacite}
\usepackage{pdflscape}
\usepackage{svg}
\usepackage{pifont}
\usepackage{url}
\usepackage{hyperref}

\theoremstyle{plain}

\theoremstyle{definition}

\theoremstyle{remark}

\begin{document}


\title{The News Comment Gap and Algorithmic Agenda Setting in Online Forums}

\author{
\name{Flora Böwing* and Patrick Gildersleve*\thanks{CONTACT: \href{mailto:}{p.gildersleve@lse.ac.uk}}\thanks{*Authors contributed equally}}
\affil{London School of Economics and Political Science, Houghton Street, London, WC2A 2AE, United Kingdom}}

\maketitle

\begin{abstract}
The disparity between news stories valued by journalists and those preferred by readers, known as the ``News Gap'', is well-documented. However, the difference in expectations regarding news related user-generated content is less studied. Comment sections, hosted by news websites, are popular venues for reader engagement, yet still subject to editorial decisions. It is thus important to understand journalist vs reader comment preferences and how these are served by various comment ranking algorithms that represent discussions differently. We analyse 1.2 million comments from Austrian newspaper Der Standard to understand the ``News Comment Gap'' and the effects of different ranking algorithms. We find that journalists prefer positive, timely, complex, direct responses, while readers favour comments similar to article content from elite authors. We introduce the versatile Feature-Oriented Ranking Utility Metric (FORUM) to assess the impact of different ranking algorithms and find dramatic differences in how they prioritise the display of comments by sentiment, topical relevance, lexical diversity, and readability. Journalists can exert substantial influence over the discourse through both curatorial and algorithmic means. Understanding these choices' implications is vital in fostering engaging and civil discussions while aligning with journalistic objectives, especially given the increasing legal scrutiny and societal importance of online discourse.
\end{abstract}


\section{Introduction} \label{introduction}

Active public debate is vital for a stable democratic society as it fosters informed citizenry and accountability for those in power. In the digital age, much of this discourse occurs online. Online platforms allow broad participation but face challenges like uncivil postings and algorithms that may reinforce existing beliefs. Hosting high-quality public debate online requires content moderation and well-designed algorithms to keep users engaged whilst avoiding strong filter bubble or echo chamber effects. Among such online platforms, news media user forums are a promising venue for public debate, acting as modern \emph{letters to the editor}. They enable engagement between journalists and readers, providing diverse perspectives. However, news organisations' interests may differ from readers', influenced by competition and (declining) profit margins, affecting the curation and presentation of both news content and user-generated content (UGC). The difference between what journalists deem worth publishing and what readers consider worth consuming, the so-called \emph{news gap}, has been thoroughly researched. However, there is only limited work on the difference between what journalists expect from UGC, and what users expect from it. This paper aims to analyse this \emph{news comment gap}. Firstly, we ask:

\begin{itemize}
    \item[\textbf{RQ1}] How do the preferences of readers and journalists differ in news comment forums?
    \begin{itemize}
        \item[\textbf{1.1}] What are the characteristics of comments that are popular/unpopular with readers?
        \item[\textbf{1.2}] What are the characteristics of comments that are selected as Editors' Picks?
        \item[\textbf{1.3}] How large is the gap between readers' and journalists' comment preferences?
    \end{itemize}
\end{itemize}

Considering the size of the discussions and users' limited time and attention, comment ranking policies are a crucial part of platform design. Comments can be ranked based on metrics reflecting both journalists' and readers' preferences, influencing which comments are prioritised. This control over comment display impacts the representation of news stories, extending agenda-setting power. We must then ask:

\begin{itemize}
    \item[\textbf{RQ2}] What are the consequences of different comment ranking algorithms on perceived discussion quality?
    \begin{itemize}
        \item[\textbf{2.1}] What is the effect of user-based ranking on perceived discussion quality?
        \item[\textbf{2.2}] What is the effect of editors' preferences via pinned comments on perceived discussion quality?
        \item[\textbf{2.3}] What is the effect of discussion structure, i.e. reply trees, on perceived discussion quality?
    \end{itemize}
\end{itemize}

The dataset underpinning this paper has been gathered from one of the largest German-speaking news media online forums, hosted by Austrian newspaper \textit{Der Standard}. Textual and contextual features of the 1m+ comments are used to model preferences through ``Editors' Picks'' and user votes through regression and classification models. Furthermore, we introduce a novel approach to measure the performance of different comment ranking algorithms, thereby showing which policies favour what kind of content.

\section{Related work} \label{related-work}

\subsection{Significance of online debate for democracy} \label{significance-of-online-debate-for-democracy}

Political discourse is a key pillar of civic engagement in a democracy. Journalism as the `fourth estate' has traditionally played a large role by informing and interacting with the public \citep{Tichenor1980}. More recently, audience participation has emerged in journalism, opening up the traditional `one-to-many' approach, enabled by the Internet \citep{Rogers2009}. Notably, comment sections underneath articles on newspaper websites as successors to ``letters to the editor'' have gained popularity \citep{Borton2013}. Such a mindset of citizen participation may help counter recent tendencies of democratic backsliding \citep{ManosevitchTenenboim2017} by fostering productive debate, but also has the potential to facilitate polarisation if platform design and moderation are poorly implemented.

Online news article comment sections emerged in the early 2000s, seen as a significant advancement in enhancing journalism's democratic character \citep{Diakopoulos2011}. Indeed, user comments in news media do have a significant impact on the on- and offline world, with them affecting perceptions of article credibility \citep{Waddell2017} and bias in media depictions of presidential candidates \citep{Houston2011}.

A comparable venue of public discourse is social media. News organisations often post the same content on their own websites and their Facebook and X/Twitter accounts, prompting comments on all platforms. News website comments differ from social media comments, even if the articles are identical. News website comments display superior deliberative quality and stay more on-topic \citep{Rowe2015, He2019}. News website comments are also more likely to suggest alternative solutions to issues, engage more with other users, and tend to contain more questions and external resources \citep{Rowe2015}. \citet{He2019} found that news platform comments tend to be less ideologically balanced than Facebook comments, while \citet{Rowe2015} found the opposite. Comments on platforms such as Facebook are expected to be much more ``socially mediated" by friends compared to comments on news websites. 

Ideally, comment sections serve as a communal public sphere for engaging with societal issues, sharing viewpoints, and fostering democratic discourse \citep{Wang2021}. In reality, issues like information overload, incivility, and institutional and legal constraints pose challenges.

\subsection{Motivations behind news engagement} \label{motivation-behind-reading-and-commenting-on-the-news}

Research has thoroughly examined why people read the news and what stories interest them. \emph{News values} are the criteria that journalists and news organisations use to determine the newsworthiness of an event or story, such as timeliness, proximity, impact, prominence, conflict, and shareability on social media \citep{Harcup2017}. Despite efforts to understand readers' preferences, a gap remains between the news supplied by journalists and readers' demands. Journalists often prioritise public-affairs stories, while readers prefer sports, entertainment, and crime news \citep{Boczkowski2013}. According to Boczkowski and Mitchelstein, this so-called ``news gap'' is considerable in substantial: stories considered most newsworthy by journalists contained 19 percentage points more public-affairs articles than those with the most reader views. \citet{Masullo2022} found a similar gap in COVID-19 local news coverage, with readers favouring information on local businesses' pandemic responses and fact-checking over economic news. \citet{Bright2016} explored the ``social news gap'', noting that social sharing behaviour is driven more by social status than emotion, with different topics being more or less popular to share. Online news consumption and rising competition make it easier for readers to selectively engage with stories, exacerbating the news gap.

Motivations for commenting on news differ from reading and sharing, affecting the nature of any news gap. While many read news and comment sections online, active commenters are a specific subset.  For example, \emph{derstandard.at} online readers are roughly gender-balanced (45\% female, 55\% male), but only 20\% of active commenters are female \citep{KrennPetrak2021}. Emotional reactions, especially negative ones like anger and anxiety, strongly predict commenting \citep{Diakopoulos2011, Naab2022, Ziegele2018}. Positive emotions have a more ambiguous effect, sometimes increasing willingness to comment and sometimes decreasing it \citep{MarcusNeumanMacKuen2000, Berger2012, Orehek2011}. The news factors of `controversy' and `damage' in articles also increase the likelihood of discussing and commenting on news \citep{Harber2005, Rogers2000, Weber2014, Ziegele2013, Boczkowski2012}. News values and discussion factors both influence the quantity and quality of comments \citep{Ziegele2018}.

Controversial political and social issues generate more comments due to their emotional impact, making comment sections engaging but prone to incivility and low-quality arguments. Enhancing debate quality remains challenging due to diverse stakeholder interests. Comment ranking policies can be a crucial lever for news organisations aiming to improve the nature of discussions in their comment sections.

\subsection{Quality of comments} \label{quality-of-comments}

Within journalism research, ``quality'' can be viewed through various lenses, such as from professional \citep{Diakopoulos2015} or reader-centred perspectives \citep{MeijerBijleveld2020}. For user comments, social signals like up- or down-voting \citep{Cheng2014} and commenting frequency \citep{Diakopoulos2011} play an important role when assessing quality. Other dimensions for evaluating comments include readability, coherence, argument quality, criticality, emotionality, personal experience, relevance, and novelty \citep{Diakopoulos2015, Wang2021}.

Many studies use ``wisdom of the crowds'' as ground truth, like user votes and author reputation for quality \citep{Momeni2016, Wei2016, Tan2016}. Others use external assessors to judge quality independently \citep{Momeni2016}, though this can introduce new biases. Regular text-based features alone are found to be ineffective for distinguishing quality; social interaction and argumentation features show more promise \citep{Momeni2016}. Based on \citeauthor{Wang1996}'s \citeyearpar{Wang1996} data quality framework for data consumers and its application to news media \citep{Tang2003}, we categorise the most common measurements of comment quality in Table \ref{tab:qualmethods}.

\emph{Intrinsic quality} refers to features of a comment assessed independently of context (i.e., other comments), such as textual features like sentiment scores and lexical diversity. \emph{Contextual quality} considers a comment's relation to the article and discussion, such as topical relevance to the article or the discussion reply structure. \emph{Representational quality} involves social status markers, like author reputation, and metrics like frequency of references to other users.

There is no universal metric for comment quality, but researchers have explored various aspects using qualitative and quantitative features. For news organisations, hosting UGC is risky, making it crucial to measure comment quality to manage their content effectively.

\begin{table*}[p]
\centering
\tbl{Methods applied to measure aspects of quality of comments and reviews from prior literature.}{
\begin{tabular}{p{2cm}|p{3cm}|p{5cm}|p{4cm}}
Category                      & Dimensions                  & Metrics                                                                                      & Used by                   \\ \hline
I. Intrinsic Quality          & Objectivity, \newline Emotionality                & - Sentiment score\newline - Civility score\newline - Personal experience score\newline - Subjectivity scores\newline - Conversational vs informational content & \citet{Pang2002}, \citet{Kim2022}, \citet{Diakopoulos2015}, \citet{Park2016}, \citet{Wang2021} \\ \cline{2-4}
                              & Believability / Credibility & - Perplexity of comment\newline - Entropy of comment\newline - Lexical diversity\newline - Informativeness\newline - Number of punctuation marks\newline - Part-of-speech \newline - Number of named entities & \citet{Otterbacher2009}, \citet{Wang2021}, \citet{Johansson2009}, \citet{Momeni2016} \\ \hline
II. Contextual Quality        & Relevance                   & - Word feature vector similarity\newline - Topic modeling features  & \citet{Otterbacher2009}, \citet{Park2016}, \citet{Momeni2016} \\  \cline{2-4}
                              & Appropriate Amount          & - \# words per comment\newline - \# sentences per comment & \citet{Park2016}, \citet{Schuth2007}, \citet{Otterbacher2009} \\ \cline{2-4}
                              & Timeliness                  & - \# days since article was published & \citet{Otterbacher2009}, \citet{Szabo2008} \\ \cline{2-4}
                              & Novelty, Originality        & - Comment word vector centroid score vs others & \citet{Otterbacher2009} \\ \cline{2-4}
                              & Agreement / disagreement      & - Normalised comment sentiment, SVMs & \citet{Fortuna2007} \\ \cline{2-4}
                              & Discussion level            & - Size of the comment tree\newline - Height of the comment tree\newline - Position in the tree\newline - \# replies\newline - Winning arguments & \citet{Schuth2007}, \citet{Wei2016}, \citet{Young2021} \\ \hline
III. Representational Quality & Ease of understanding       & - Characters-to-sentences ratio\newline - Words-to-sentences ratio\newline - Readability (e.g., the SMOG reading grade level) & \citet{Otterbacher2009}, \citet{Park2016} \\ \cline{2-4}
                              & Social presence             & - Tagging or thanking other users & \citet{Swan2019} \\ \cline{2-4}
                              & Author reputation           & - Registration age\newline - \# contacts\newline - \# past posts from author \newline - Author’s engagement and status in the social network & \citet{Momeni2016}, \citet{Castillo2011}, \citet{Canini2011}
\end{tabular}}
\label{tab:qualmethods}
\end{table*}

\subsection{Institutional aspects of user-generated content} \label{institutional-aspects-of-user-generated-content-in-news-organisations}

Over the past decades, news media platforms have invested in interactive elements to engage users. Qualitative research on UGC suggests it can enhance democratic participation and promote political efficacy \citep{ManosevitchTenenboim2017}. Journalists benefit from ``reciprocal journalism'', fostering relationships with readers \citep{Chen2017}. User comments facilitate analytical and social deliberation and active discussions between opposing views \citep{ManosevitchWalker2009, Kelly2005}. However, some newspapers struggle with adhering to and enforcing their participation rules \citep{Noci2012}, leading to shutting down comment sections to reduce misinformation and incivility \citep{Nelson2021}.

The EU Digital Services Act imposes stricter moderation requirements for UGC, mandating the removal of hate speech, illegal content, and misinformation, with fines up to 6\% of annual global revenue \citep{EuropeanUnion2022}. Similar legislation is being implemented and considered in the UK (Online Safety Act) and the US with challenges to Section 230. While these regulations may improve comment quality, they will also increase the resources needed for moderation. Smaller EU organisations face lower fines, reducing their risk. However, without size considerations in the UK or US, newspapers might shut down comment sections or enforce overly strict moderation, harming freedom of speech online \citep{Article192022}.

UGC offers news organisations opportunities to increase engagement and revenue through subscriptions or advertisements. However, it can also lead to incivility, and with stricter legal scrutiny and high fines for hate speech, the risk has grown. Effective moderation, combining human and automated judgement with platform design features like ranking policies, is essential.

\subsection{Information overload and ranking policies} \label{information-overload-and-ranking-policies}

The scale of online discussions encountered by users is considerably larger than that offline, leading to information overload. Users often lack the time to scan hundreds of comments, missing potentially interesting ones and sometimes ending their participation altogether \citep{Hogan2009, Levy2008, Lampe2014}. On some platforms, users view only 3\% of comments \citep{Althuniya2020}. Automated assessment and ranking are increasingly important \citep{Momeni2016}. Ideally, users see a summary of the discussion through selected comments, though this is challenging without biasing the results. Many news organisations rank comments in reverse-chronological order by default, or allow sorting by user up/downvotes. Some run an \emph{Editors' Picks} section, where selected comments are made prominently visible to readers. 

Since most users don't change the default sorting \citep{Lampe2014}, the default policy choice has a significant impact. Comment sorting policies can improve comment quality, boost user participation, and foster constructive discussion \citep{Diakopoulos2011}. For instance, \citet{Wang2021} found that New York Times users whose comments were selected as Editors' Picks subsequently wrote higher-quality comments. Exposure to thoughtful comments encourages users to write more thoughtfully \citep{Sukumaran2011} and increases their willingness to participate in discussions \citep{Han2015}.

Ranking policies based on the ``wisdom of the crowds'' (e.g., user votes) can create social pressure, leading users to comment for upvotes rather than content quality \citep{Heiss2019, MarcusNeumanMacKuen2000}. Disapproving responses can discourage even civil commenters, prompting them to withdraw \citep{Naab2022}. Naab suggests that social sanctioning can suppress certain voices, requiring attentive moderation and user support to prevent \citep{Ziegele2020}. Social feedback can heavily influence discussions, potentially sanctioning divergent opinions, and news organisations' political leanings may amplify this pressure.

In online forums, comment ranking influences which comments users read as well as write. Editor-picked comments provide a debate overview but may reflect undisclosed editorial policies. Community-based ranking can sanction incivility but may discourage minority opinions. This creates tension between audience participation and journalistic gatekeeping \citep{Chen2017, Diakopoulos2015}. Users' and editors' judgements often differ; only 17.2\% of New York Times comments made both the readers' and Editors' Picks lists. Readers favour direct, confrontational, and aligned comments, while editors chose more articulate, conciliatory, and diverse ones \citep{JuarezMiro2022}.

Building on \citeauthor{JuarezMiro2022}'s (\citeyear{JuarezMiro2022}) work, we examine the ``news comment gap'' by using computational methods to analyse the characteristics that correlate with reader votes and Editors' Picks. We compare the discussion qualities under different ranking policies, and analyse the impact of editorial features like Editors' Picks and structural features like reply-trees. Existing research on news comments has predominantly focused on English-speaking platforms, as well as specifically on hate-speech \citep{Reimer2021}. This study explores comment quality and ranking algorithms in the largest German-speaking news forum, \emph{derstandard.at}.

\section{Data and Methodology} \label{data-and-methodology}

\subsection{Data} \label{data}

\begin{figure}
    \centering
    \includegraphics[]{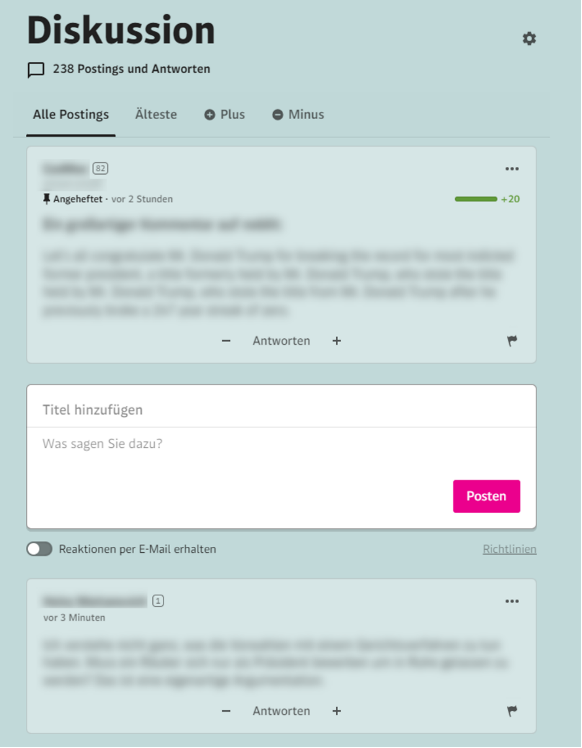}
    \caption{Exemplary screenshot of the comment section underneath an article. Four tabs enable users to sort comments in different ways (``Alle postings" = ``all postings"/reverse-chronological/newest first, ``Älteste" = ``oldest first"/chronological, ``Plus" = most upvotes, ``Minus" = most downvotes). In this case, one pinned comment is shown at the very top, followed by the input box and the first comment of the general sorting. Upvotes and downvotes are shown in the top-right corner of each comment in green and red, respectively. A user can reply to a comment by clicking ``Antworten", upvote by clicking ``+" and downvote by clicking ``-". Usernames and text redacted for privacy reasons.}
    \label{fig:enter-label}
\end{figure}

\textit{Der Standard} is one of Austria's two most-read ``quality-newspapers'' (``qualitätszeitung'', roughly equivalent to a ``broadsheet''), hosting one of the largest German-speaking online forums globally with over 20 million comments by 73,000 active users annually \citep{Burger2022}. The online version has no paywall and readers can freely access all articles, read all comments, and sort comments by time published and user votes. To write comments or vote on comments, readers need to register (pseudonymously, if desired). The interface, in Figure \ref{fig:enter-label}, offers features to change the ranking policy (\emph{newest, oldest, most positive}, and \emph{most negative}) and to hide or show replies of each level individually. Pinned posts (or ``Editors' Picks'') are comments deemed especially valuable by  editors, community management, or community guides\footnote{``Editors'' here includes the journalist(s) who wrote the article. ``Community management'' are \textit{Der Standard} employees in charge of a small number of volunteer ``Community guides'' who also assist with content moderation \citep{DerStandard2021}}. They are shown at the top of the discussion and marked as \emph{pinned} (``\emph{Angeheftet}'').

The dataset used in this paper was scraped using RSelenium \citep{RSelenium} from \emph{derstandard.at}. The dataset contains 5,874 articles published in September and October 2022 and their comments (1.2 million). Summary statistics are presented in Table \ref{tab:summary-stats}. There are on average 204.8 comments per article discussion. The discussion sizes is a long-tailed distribution with many articles having relatively few comments (median = 79), and a small number of discussions being very large. The average comment length is 27.6 words or 184.3 characters. 31.6\% of comments are root comments, 32.2\% are first-level replies, 17.0\% are second-level replies, and the remaining 19.2\% are third-level replies or higher.

The comment-level dependent variables of interest are \emph{Upvotes} and \emph{Downvotes} measuring user votes, as well as \emph{Editors' Pick} measuring whether a comment has been selected by a journalist to be pinned at the top of a comment section. The vast majority of comments have at least one user up/down vote (mean = 6.7, median = 3.0). There are many more upvotes than downvotes, with an average of 5.4 upvotes and 1.4 downvotes per comment. 0.5\% of all comments are pinned. Pinned comments are only in larger discussions (mean no. of comments/discussion = 537.5), which make up about 25\% of the dataset. The remaining 75\% of discussions are significantly smaller. In other words, most comments are part of discussions with pinned comments.

\begin{table}[h]
    \centering
    \tbl{Summary statistics for article comment sections.}{
    \begin{tabular}{l|r|r}
    Metric & w/ Editors' Picks & w/o Editors' Picks \\
    \hline
    Number of discussions & 1,511 & 4,363 \\
    Total comments & 812,088 & 391,307 \\
    Avg comments per discussion & 537.45 & 89.69 \\
    Avg user votes per comment & 7.0 & 4.45 \\
    \end{tabular}}
    \label{tab:summary-stats}
\end{table}

\textit{Der Standard} uses various moderation techniques. Comments are monitored by employees (community management) and volunteers (community guides), aided by machine learning software \citep{DerStandard2022}. In 2021, only 2.4\% of comments were deleted for violating community standards \citep{Burger2022}. Nevertheless, it is worth considering that moderation rules and execution affect the subset of comments published and therefore the discussion quality shown through different ranking algorithms.

\subsection{Feature Selection and Models} \label{feature-selection-and-models}

\subsubsection{Features}
We selected 19 comment-level features of interest based on metrics used in previous research (see Table \ref{tab:qualmethods} in the \emph{Related Work} section). These are summarised in Table \ref{tab:features}. We selected these relatively off-the-shelf features for consistency and comparability with prior work and to illustrate fundamental differences between journalists and readers, as well as taking into account limited computational resources. We also add article-level controls for the level of article engagement (number of comments, mean up/downvotes in discussion) and article genre variables to account for differing natures of discussion by genre. Furthermore, in Section \ref{comparing-ranking-algorithms}, a subset of these features is used to highlight how even simple ranking algorithms can significantly alter the kinds of comments prioritised for display to users. 

Some comments only contain a few words or a link, leading to missing values in the cosine similarity score, lexical diversity score, and readability score. In the former two cases, we set the missing values to zero, as short comments are assumed to be irrelevant and too short to be lexically diverse. For the readability score, we set the missing values to the minimum, as these are the shortest and simplest comments. To assist with model convergence/fit, and for appropriate comparison between features, we log transform ($x' = \log(1+x)$) the long-tailed explanatory variables (Author follower count, Level in tree, Size of tree, Height of tree, Hours since article, Num punctuation, Num replies, Num Upvotes, Num Downvotes, Num of comments, Mean upvotes in discussion, and Mean downvotes in discussion). For the same reasons, we standardise all numeric explanatory variables to $\mu=0$, $\sigma=1$. To answer RQ2, we selected a subset of 4 of the 19 features: a \textit{compound} sentiment score ($= \text{positive sentiment} - \text{negative sentiment}$), readability, topical relevance, and lexical diversity.

\subsubsection{Regression Analysis} \label{regression-analysis}

To address RQ1, we use a subset of the features that only factor in information available at the time of comment posting. We first fit a number of regression models to discover associations between comment features and user votes (RQ1.1) / Editors' Picks (RQ1.2). We use negative binomial regressions for Upvotes and Downvotes and a binary logistic regression model for Editors' Pick status (in this case, only root comments are eligible to be selected, so we drop the `is root comment' and `comment level in tree' variables). We use 80\% of the data (962,716 randomised comments) to fit the models and test them using the remaining 20\%. We compare coefficients from the models to understand journalist and readers' preferences. After answering RQ1, we build predictive models for upvotes and Editors' Picks using all of the features that we use as mock ranking algorithms for RQ2 (further details in Appendix \ref{rank-models}). These models take all available features and produce a ranking of comments based on predicted upvotes / Editors' Picks. We do this with negative binomial and logistic regressions, as well as applying a more complex statistical learning approach with XGBoost to predict user votes / Editors' Picks.

\subsubsection{XGBoost} \label{xgboost}

We use the XGBoost library (eXtreme Gradient Boosting) \citep{chen2016xgboost} for regression on upvotes and classification for Editors' Picks. We used successive random search and narrower grid search to tune hyperparameters for both models. For regression, we chose root mean squared logarithmic error loss (for appropriate comparison to the negative binomial regressions), and for classification we chose cross-entropy loss.

\subsubsection{Measuring the comment gap} \label{measuring-the-comment-gap}

The size of the news comment gap between user votes and Editors' Pick preferences (RQ1.3) can be calculated through the Jaccard coefficient, following \citet{JuarezMiro2022}. We consider only the largest discussions where comments are pinned (N = 1,511). For each discussion, we calculate the overlap of \emph{p} pinned Editors' Pick comments with the \emph{p} most-liked comments in terms of upvotes and relative votes.

\begin{landscape}
\begin{table}[t]
    \centering
    \tbl{Features selected to model reader and editor preferences through user votes / Editors' Picks.}{
    \begin{tabular}{l|l|l|l}
    Quality & Category & Feature & Comment \\ \hline
    Intrinsic quality & Objectivity/emotionality & Sentiment score (positive) & Sentiment analysis via German text BERT model \citep{Guhr2020} \\ 
     &  & Sentiment score (negative) &  \\ \cline{2-4}
     & Credibility & Lexical diversity & Carroll’s Corrected Type-Token-Ratio \citep{Benoit2018} \\ 
     &  & Num punctuation & Number of punctuation marks, e.g. “!!!” \\ \cline{2-4}
     & Appropriate amount & Num sentences & Number of sentences \\  \cline{2-4}
     & Ease of understanding & Readability &  Adapted SMOG score for German text \citep{Benoit2018, McLaughlin1969} \\ \hline
    Contextual quality & Relevance & Topical similarity & TF-IDF cosine similarity of the comment to the article \\  \cline{2-4}
     & Timeliness & Hours since article & Hours passed since the article was published \\  \cline{2-4}
     & Discussion level & Is pinned & Whether a comment has been pinned as an Editors' Pick \\ 
     &  & Is leaf comment & Whether a comment has replies (no leaf) or not (leaf) \\ 
     &  & Is root comment & Whether a comment is the first one in a reply tree \\ 
     &  & Size of tree & Total number of comments in the reply tree \\ 
     &  & Height of tree & Total number of levels in the reply tree \\ 
     &  & Level in tree & Level of a comment inside a reply tree \\ \hline
    Representational quality & Interaction features & Num replies & Total number of direct and indirect replies to a comment \\  
     &  & Num Upvotes & Total number of Upvotes a comment receives \\  
     &  & Num Downvotes & Total number of Downvotes a comment receives \\  \cline{2-4}
     & Social presence & Second person & Whether a comment contains a 2\textsuperscript{nd} person pronoun (“Du”/“Sie”) \\ 
     &  & Author follower count & The number of ``followers'' a comment author has \\ 
    \end{tabular}}
    \label{tab:features}
\end{table}
\end{landscape}

\subsection{Evaluating Comment Ranking Algorithms}

\subsubsection{Constructing Ranking Algorithms} \label{constructing-ranking-algorithms}

In order to answer RQ2.1-3, we first define different possible comment ranking policies. Each comment ranking algorithm, or ``sorting policy'', is made up of three policy elements: primary ordering, Editors' Pick status, and reply structure. Primary ordering is the main way in which comments are ranked relative to each other (e.g. in order of upvotes, time posted, etc.). We implement 11 different primary orderings. Editors' Pick status determines whether selected comments are pinned to the top of the discussion or treated as regular comments to be ranked. There are three reply structure options: hiding replies, showing full reply trees attached to their root comment, and showing replies ``loose''--unattached to their root comment. When full reply trees are shown, the root comments are sorted by the primary ordering, but the reply trees retain sorting by comment time and discussion structure. In total, there are $11 \times 2 \times 3 = 66$ different combinations of policy elements that may make up a ranking algorithm, summarised in Table \ref{tab:sortcats}.

\begin{table}[h]
\centering
\tbl{Ranking policy elements organised by category. We add predictive ranking algorithms in addition to the simple measures.}{
\begin{tabular}{l|l}
Category                & Ranking options                                        \\ \hline
User-based (primary ordering) (RQ2.1)  & \begin{tabular}[c]{@{}p{5cm}@{}} - Most upvotes first \\ - Most relative votes first (upvotes – downvotes) \\ - Least downvotes first \\ - Most downvotes first \\ - Chronological \\ - Reverse Chronological  \\- Random \\ - \textit{Predicted upvotes (regression)} \\ - \textit{Predicted upvotes (random forest)} \\ - \textit{Predicted Editors' Picks (regression)} \\ - \textit{Predicted Editors' Picks (random forest)} \end{tabular} \\ \hline
Editor-based (RQ2.2)    & \begin{tabular}[c]{@{}p{5cm}@{}} - Pinned Editors' Picks first \\ - Unpinned Editors' Picks \end{tabular} \\ \hline
Structure-based (RQ2.3)  & \begin{tabular}[c]{@{}p{5cm}@{}} - All replies hidden \\ - Replies shown in fixed trees  \\ - All comments loose \end{tabular}               
\end{tabular}}
\label{tab:sortcats}
\end{table}

We select a subset of features to measure the performance of different sorting policies over: compound sentiment score ($= \text{positive sentiment} - \text{negative sentiment}$), readability score, lexical diversity, and topical relevance. These four meaningful metrics are shown to be associated with reader/editor comment preferences and we seek to evaluate how different sorting policies may prioritise reader/editor preferred comments. For each of the selected features, we offset the variable so that all values are non-negative (if needed).

\subsubsection{Comparing Ranking Algorithms with FORUM} \label{comparing-ranking-algorithms}

The subsequent analysis of different ranking algorithms is partly inspired by \citet{Young2021}, who compared sorting policies on a debating website with regard to \emph{winning arguments} in argumentation theory. For each discussion, we order the comments according to each ranking policy. We devise the ``\textbf{F}eature-\textbf{O}riented \textbf{R}anking \textbf{U}tility \textbf{M}etric'' (FORUM) that measures how well a ranking policy $p$ surfaces the best comments in a discussion according to a feature $f$. The metric compares how the ordering of comments $o^p$ under policy $p$ compares to the best-possible ordering ($o^b$ - descending order of score $f$), worst-possible ordering ($o^w$ - ascending order of score $f$), and the expected random baseline. We can use FORUM to evaluate how well a full discussion of $N$ comments is ordered, or the first $n<N$ comments returned. Here we describe how FORUM is calculated for a ranking policy $p$ with feature $f$ on a particular discussion on a news article.

\begin{figure}
    \centering
    \includegraphics[]{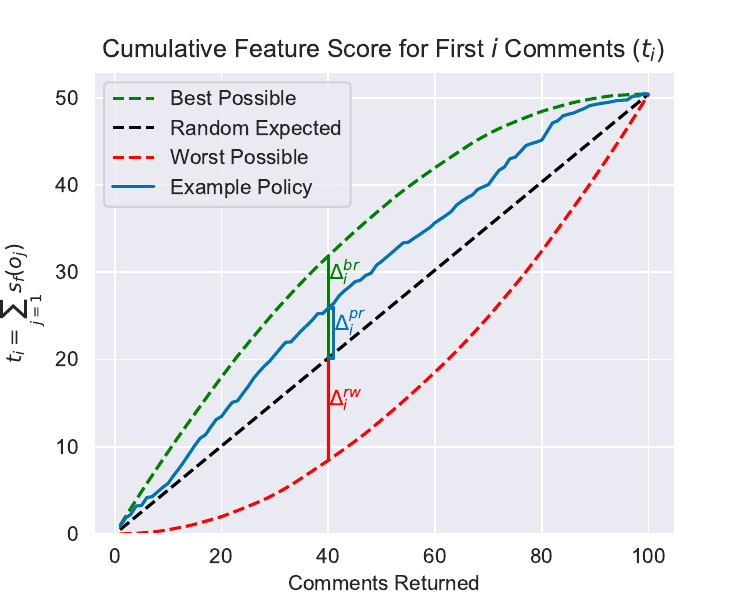}
    \caption{Cumulative feature score for the first $i$ comments presented to readers when comments are ranked by best possible (descending order), worst possible (ascending order), random, and example ranking policies.}
    \label{fig:feature_score}    
\end{figure}

If a user is shown $i$ comments from a discussion, they will be exposed to $t_i = \sum_{j=1}^{i} s_f(o_j)$ cumulative feature score, where $o_j$ is the comment at position $j$ in ordering $o$, and $s_f(o_j)$ is the feature score of said comment. The expected random baseline cumulative score is defined by $t_i^r = iT_f/N$, where $N$ is the total number of comments and $T_f$ is the total score of the $N$ comments. We measure the difference in cumulative scores to the random baseline for the best-possible ordering, worst-possible ordering, and policy $p$: $\Delta_i^{br} = t_i^b - t_i^r$, $\Delta_i^{rw} = t_i^r - t_i^w$, and $\Delta_i^{pr} = t_i^p - t_i^r$ (note that $\Delta_i^{br}, \Delta_i^{rw} \ge 0$).  A graphical interpretation of this is shown in Figure \ref{fig:feature_score}. If a policy tends to return higher scored comments earlier than expected at random, the line tracking cumulative score will fall above the diagonal baseline. If the policy tends to return higher scoring comments later than expected at random, then it will fall below the baseline. Then, to compare performance relative to the best/worst possible ordering in showing $i$ comments, the ``normalised policy delta'' is:

\begin{equation}
    \Gamma^p_i = \frac{\Delta_i^{pr}}{(\Delta_i^{br})^{H(\Delta_i^{pr})} (\Delta_i^{rw})^{1-H(\Delta_i^{pr})}},
\end{equation}

where $H(\Delta_i^{pr})$ is the Heaviside step function ($= 1$ for $\Delta_i^{pr} \ge 0$, $= 0$ for $\Delta_i^{pr} < 0$), allowing for differential normalisation when $\Delta_i^{pr}$ is positive or negative. Policies that return many of the highest scoring comments in the first $i$ comments achieve $\Gamma^p_i \approx 1$, while those returning many of the lowest scoring comments achieve $\Gamma^p_i \approx -1$, and those no better/worse than random achieve $\Gamma^p_i \approx 0$.

The normalised policy delta measure, $\Gamma^p_i$, can be used to measure the total quality of the first $i$ comments returned, compared to best/worst/random. However, it \emph{does not} measure how well the $i$ comments themselves returned are ranked. To measure how well a policy $p$ ranks up to $n$ comments, we must take the \emph{average} normalised policy delta for all $i$ up to $n$:

\begin{equation}
    \Phi_n^p = \frac{1}{n}\sum_{i=1}^{n}\Gamma^p_i = \frac{1}{n}\sum_{i=1}^{n}\frac{\Delta_i^{pr}}{(\Delta_i^{br})^{H(\Delta_i^{pr})} (\Delta_i^{rw})^{1-H(\Delta_i^{pr})}}.
\end{equation}

\begin{figure}
    \centering
    \includegraphics[]{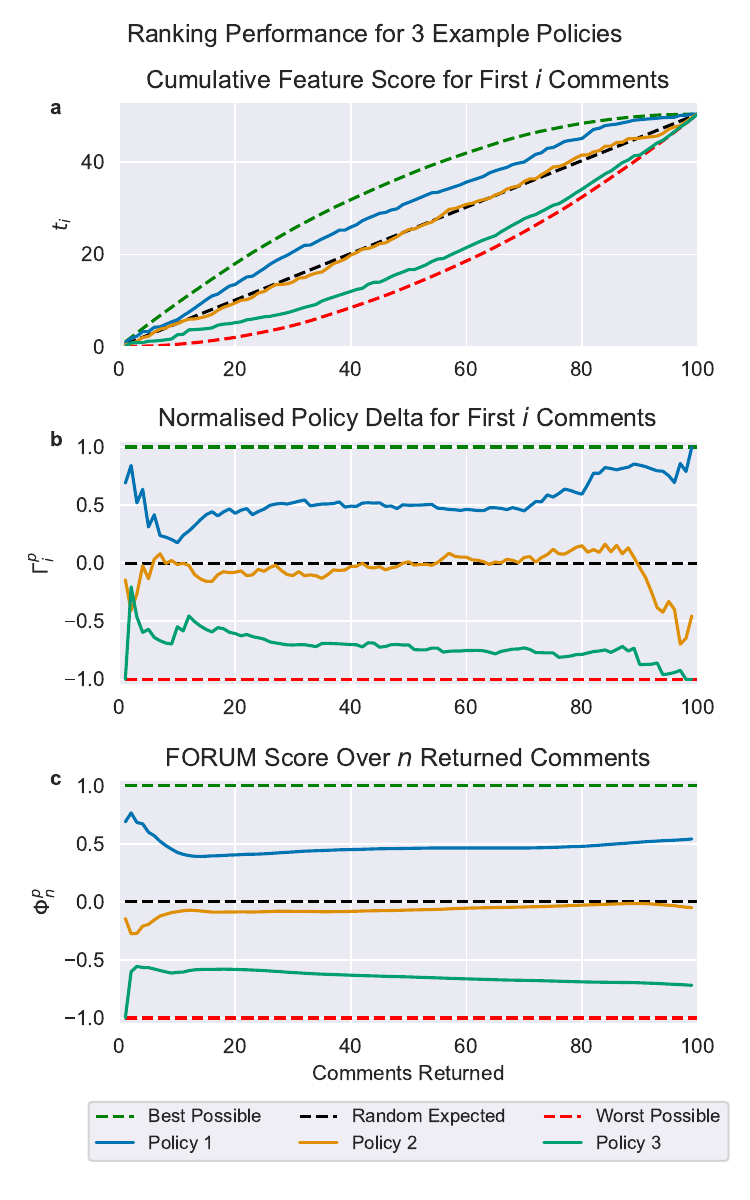}
    \caption{The process of calculating FORUM score over n comments for three example policies. First the cumulative feature scores for the first $i$ comments returned by each policy are calculated (a). The performance above/below random is then normalised to the best/worst possible policies (b). Finally, the normalised scores are averaged to determine the FORUM score for how well the policy ranks up to $n$ comments (c). Policy 1 performs better than random, policy 2 performs close to random, and policy 3 performs worse than random.}
    \label{fig:sorting_performance}    
\end{figure}

$\Phi_n^p$ is then our \textbf{F}eature-\textbf{O}riented \textbf{R}anking \textbf{U}tility \textbf{M}etric (FORUM). Here, the ordering of the $n$ comments returned is considered, since we average normalised policy delta for the first $1, 2, 3, ..., n$ comments. Even if two policies return the same $n$ comments (same $\Gamma^p_n$), the one that returns higher scoring comments earlier will amass cumulative comment score, $t_i^p$, earlier, and will have higher $\Gamma^p_i$ for $i < n$. Thus, FORUM for $n$ comments, $\Phi_n^p$, will be larger. The full process for three example policies is shown in Figure \ref{fig:sorting_performance}.

Note that $\Gamma^p_N$ is undefined due to division by zero. Intuitively, there is only one way to return a set of $N$ out of $N$ total comments (i.e., all of them), so a relative performance score does not make sense. Relatedly, this means to score a ranking policy over a full discussion of $N$ comments we take $\Phi_{N-1}^p$---scoring a policy for how it ranks a full discussion is equivalent to scoring it for $N-1$ comments, since there is only one position for the final comment to go.

In cases where a ranking policy removes comments from immediate view, such as the ``hide replies'' functionality on \textit{Der Standard}, we choose to interpolate a straight line from the cumulative feature score of the final visible comment up to the total cumulative feature score $T_f$ in the process of calculating $\Phi_n$. This allows comparison between ranking policies over a different number and set of comments. Practically, this is akin to a situation where any one of the hidden comments is equally likely to be viewed next by a reader (if they choose to click any to expand).

The FORUM score is bounded $-1 \le \Phi^p_n \le 1$ and readily interpretable. Policies close to best-possible ordering achieve $\Phi^p_n \approx 1$, those close to worst-possible ordering achieve $\Phi^p_n \approx -1$, and those no better/worse than random achieve $\Phi^p_n \approx 0$. The measure allows comparison between policies for fixed $n$ (e.g. how well do policies $p_\alpha$, $p_\beta$ do at returning the first 10 comments in a discussion) and differing $n$ (e.g. how well do policies $p_\alpha$, $p_\beta$ do at ranking the first $10, 20, N$ comments). The measure is also normalised such that one can compare policy performance over different comment sections, even if average feature scores between them differ---say the topic of one comment section lends itself to comments with higher readability score, compared to another---(e.g. how well does policy $p_\alpha$ do at ranking the first $n$ comments in discussions 1 and 2, or how well does it do at ranking the full discussions of length $N_1, N_2$).

Despite these normalisations, the scores/weights of each comment in a discussion are still factored in when evaluating policy performance, beyond just considering the rank order of comments returned (like a simple rank correlation measure would do). $\Phi^p_n$ is intuitively similar to an area-under-receiver-operating-characteristic measure that takes into account item weight/score and is normalised to best/worst-possible ranking performance (since sorting is a much more constrained problem compared to independent binary classification). A somewhat comparable metric would be Normalised Discounted Cumulative Gain (nDCG) \citep{Wang2013}, however, this measure is not readily interpretable against worst-possible or random baseline ordering, and is particularly sensitive to the scale of the `relevance' score, often taken as binary.

\subsubsection{Modelling FORUM} \label{modelling-forum}

Each of the 66 ranking policies is applied to every article comment section, with FORUM score recorded for each feature after 10 comments ($\Phi_{10}$) and over the full discussion ($\Phi_N$). The full discussion is used to assess overall performance of each policy. The first 10 comments are used since regular users are more likely to only read the first comments displayed, and policy performance over these ranges is not necessarily correlated. In practice, this measure over the top ranked comments is likely a better indicator of the impact of different sorting policies on user experience. We seek to evaluate how well different comment ranking policies return comments by each feature over all of the article comment sections. For each feature, we perform a beta regression to evaluate the ranking performance of each policy, measured by FORUM score, and how it depends on their constituent policy elements. We perform the necessary transformation on the outcome variable ($\Phi_n$) from range ($-1$,$1$) to ($0$,$1$) by taking $\Phi'_n = (\Phi_n + 1)/2$.

For the explanatory variables in each regression, we create 10 dummies for each user-based policy element (with random ordering as a baseline), 1 dummy for the editor-based policy element to pin Editors' Picks (with no pinned comments as baseline), and 2 dummies for structure-based policy elements (with replies shown loose as baseline). With these selections, expected FORUM score on any given feature for a random ordering, no pinned comments, replies loose policy is 0, and the alternative policies are compared against this. Each coefficient then measures the expected change in ranking performance from baseline associated with employing each policy element. In total, this results in 8 regressions, one for each of the 4 target features and 2 discussion lengths, all with 13 main explanatory variables. We perform model selection (Appendix \ref{regression-specifications}) to decide which terms and interactions to use. In the final model, these explanatory variables are evaluated as main effects, with interaction terms between primary ordering and reply structure, and primary ordering and Editors' Pick status also evaluated. The final model terms with interactions are summarised in Table \ref{tab:reg-terms}.

\subsubsection{Code, Data, and Ethics} \label{ethics}

The project's repository is hosted at \href{https://github.com/dornleiten/commentgap}{\textit{github.com/dornleiten/commentgap}}. Comment data is not published to preserve user privacy. This work was approved by university ethical review procedures (ID 179954). The dataset is based exclusively on public data and was scraped according to \emph{derstandard.at}'s terms \& conditions \citep{DerStandard2024}. The authors report there are no competing interests to declare.

\begin{landscape}
\begin{table}
\centering
\tbl{Terms in the regression models organised by policy element. Reference categories (random primary ordering, loose replies, unpinned Editors' Picks) shown in italics.}{
\begin{tabular}{l|l|l|l|l}
Primary Ordering  & Reply Structure   & Editors' Picks Status  & Primary Ordering $\times$ Reply Structure Interaction  & Primary Ordering $\times$ Editors' Picks Status Interaction \\ \hline
\begin{tabular}[c]{@{}l@{}}\textit{Random}\\ Upvotes\\ Relative Votes\\ Downvotes\\ Rev. Downvotes\\ Chronological\\ Rev. Chronological\\ Pred. Editors' Picks (LR)\\ Pred. Editors' Picks (RF)\\ Pred. Upvotes (NBR)\\ Pred. Upvotes (RF)\end{tabular} & \begin{tabular}[c]{@{}l@{}}\textit{Loose}\\ Hidden\\ Fixed in Trees\end{tabular} & \begin{tabular}[c]{@{}l@{}}\textit{Unpinned}\\ Pinned at top\end{tabular} &  $\left( \begin{array}{l} \text{\textit{Random}} \\ \text{Upvotes} \\ \text{Relative Votes} \\ \text{Downvotes} \\ \text{Rev. Downvotes} \\ \text{Chronological} \\ \text{Rev. Chronological} \\ \text{Pred. Editors' Picks (LR)} \\ \text{Pred. Editors' Picks (RF)} \\ \text{Pred. Upvotes (NBR)} \\ \text{Pred. Upvotes (RF)} \end{array} \right) \times \left( \begin{array}{l} \text{\textit{Loose}} \\ \text{Hidden} \\ \text{Fixed in Trees} \end{array} \right)$  &  $\left( \begin{array}{l} \text{\textit{Random}} \\ \text{Upvotes} \\ \text{Relative Votes} \\ \text{Downvotes} \\ \text{Rev. Downvotes} \\ \text{Chronological} \\ \text{Rev. Chronological} \\ \text{Pred. Editors' Picks (LR)} \\ \text{Pred. Editors' Picks (RF)} \\ \text{Pred. Upvotes (NBR)} \\ \text{Pred. Upvotes (RF)} \end{array} \right) \times \left( \begin{array}{l} \text{\textit{Unpinned}} \\ \text{Pinned} \end{array} \right)$
\end{tabular}}
\label{tab:reg-terms}
\end{table}
\end{landscape}

\section{Results} \label{results}

\subsection{The comment gap (RQ1)} \label{the-comment-gap-rq1}

\begin{figure*}
    \centering
    \includegraphics[width=\textwidth]{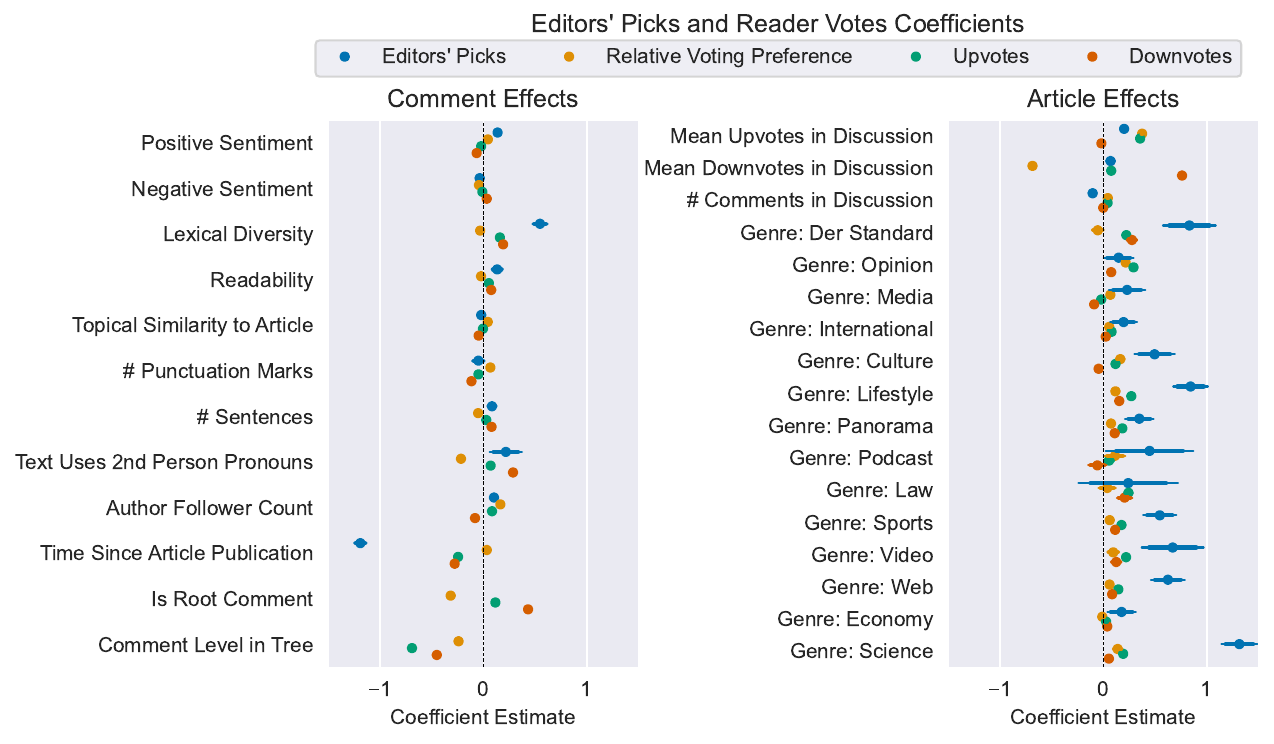}
    \caption{Regression coefficients showing journalist and reader preferences for different comment characteristics (equivalent to model coefficients in the case of Editors' Picks, Upvotes, and Downvotes). 95\% and 99\% confidence intervals indicated (many obscured by point size). Full data in Tables \ref{tab:pinvote-regressions} and \ref{tab:logodds-regression}.}
    \label{fig:pinvotes-regression}
\end{figure*}

First, we examine the results from logistic and negative binomial regression models for Editors' Picks, Upvotes, and Downvotes to understand their relationship with various comment features. All continuous variables are standardised and log-transformed where necessary, prior to model fitting as detailed in Section \ref{feature-selection-and-models}. Model summaries with their coefficients are provided in Table \ref{tab:pinvote-regressions}. Since users are able to both Upvote and Downvote comments, and that one would expect both of these factors to be associated with the level of exposure/attention a comment receives, we also compute a score to assess the relative preferences of readers by the coefficients for upvoting/downvoting a comment. For each feature, the Relative Voting Preference (RVP) is defined by 
\begin{equation}
    \text{Relative Voting Preference} = \beta_{\text{Upvote}} - \beta_{\text{Downvote}}.
\end{equation}
Taking $e^{\text{RVP}}$ then describes the factor by which the count of Upvotes on a comment increases relative to Downvotes when there is a one unit increase in the explanatory variable. Similarly, for each feature we compute the Comment Gap (CG) to assess the difference between journalist and reader preferences 
\begin{equation}
    \text{Comment Gap} = \beta_{\text{Editors' Pick}} - (\beta_{\text{Upvote}} - \beta_{\text{Downvote}}).
\end{equation}
Taking $e^{\text{CG}}$ then describes the factor by which the odds of a comment being selected as an Editors' Pick increase compared to the relative Upvotes vs Downvotes when there is a one unit increase in the explanatory variable. These are presented in Table \ref{tab:logodds-regression}. In addition to the model summary tables, we visualise the comment-level and article-level (standardised) coefficients for each model in Figures \ref{fig:pinvotes-regression} and \ref{fig:commentgap-logodds}.

\subsubsection{User votes} \label{i.-user-votes}

The following findings give an indication about which characteristics of comments are associated with Upvotes/Downvotes (RQ1.1). We may take the negative binomial regression coefficients from Table \ref{tab:pinvote-regressions} and exponentiate them to get the rate ratios that describe the factor by which Upvotes/Downvotes increase with a one unit increase in the explanatory variable.

More instructive, however, are Relative Voting Preferences in Table \ref{tab:logodds-regression}. These offer a better assessment for the relative preferences of readers. Exponentiating the RVPs describes the factor by which the count of Upvotes on a comment increases relative to Downvotes when there is a one unit increase in the explanatory variable. Here, we find that readers prefer comments with higher Positive Sentiment and lower Negative Sentiment. Increasing Lexical Diversity, Readability score, Number of Sentences decreases the Relative Voting Preference, indicating these factors more strongly increase the number of Downvotes a comment receives. Readers thus prefer simpler, shorter comments. Readers prefer comments with greater Topical Similarity to the Article and with more Punctuation Marks. Readers also prefer comments from those with higher Author Follower Count, indicating a quality or reputational effect. Increased Time Since Article Publication decreases the rate ratio of a comment being upvoted less than it being downvoted, so readers relatively prefer later comments. On structural factors, a comment being a Root Comment increases the rate ratio of it receiving Downvotes more than Upvotes. In addition, as the Comment Level in Tree increases, the rate ratio of Downvotes to Upvotes increase further. This is a slightly odd result, but can be explained by the frequency of comments at each level. There are fewer comments at each increasing level, but they are relatively more likely to attract Downvotes compared to Upvotes (likely due to long arguments). The majority of non-root comments (direct and secondary replies) have greater rate ratios of receiving Upvotes compared to Downvotes. At the article level, trivially, the Mean Upvotes and Mean Downvotes in Discussion are associated with increased rate ratios of upvoting and downvoting respectively. Readers prefer comments in discussions where the Number of Comments is higher---popular articles attract more positive engagement. On genre, readers are relatively more inclined to Upvote than Downvote articles in most genres, compared to the baseline Domestic category, the strongest of which being the Opinion genre. The exceptions being Women's Issues (negative Relative Voting Preference, weakly significant), Law and Economy (no significant difference compared to Domestic).

\begin{table}
\centering
\tbl{Logistic and negative binomial regression model summaries with coefficients for a comment being selected as an Editors' Pick or receiving Up/Downvotes. Predictive performance (F1) for Editors' Picks is low due to the highly imbalanced data, limited features, and simple model used. Note that coefficients in the Editors' Picks model for the structural factors are not fit here, since only root comments are eligible to be selected by journalists. ***: $p < 0.001$, **: $p < 0.01$, *: $p < 0.05$.}{
\begin{tabular}{l|ccc}
Feature & Editors' Picks & Upvotes & Downvotes\\ \hline
Intercept & -5.470*** & 1.098*** & -0.405*** \\
Positive Sentiment & 0.137*** & -0.021*** & -0.065*** \\
Negative Sentiment & -0.036* & -0.010*** & 0.033*** \\
Lexical Diversity & 0.548*** & 0.160*** & 0.191*** \\
Readability & 0.134*** & 0.053*** & 0.076*** \\
Topical Similarity to Article & -0.021 & -0.004** & -0.046*** \\
\# Punctuation Marks & -0.048* & -0.048*** & -0.115*** \\
\# Sentences & 0.083*** & 0.028*** & 0.079*** \\
Text Uses 2nd Person Pronouns & 0.217*** & 0.070*** & 0.287*** \\
Author Follower Count & 0.103*** & 0.084*** & -0.081*** \\
Time Since Article Publication & -1.191*** & -0.245*** & -0.278*** \\
Is Root Comment & - & 0.116*** & 0.433*** \\
Comment Level in Tree & - & -0.691*** & -0.451*** \\
Mean Upvotes in Discussion & 0.201*** & 0.356*** & -0.019*** \\
Mean Downvotes in Discussion & 0.069*** & 0.076*** & 0.762*** \\
\# Comments in Discussion & -0.104*** & 0.041*** & -0.002 \\
Genre: Women's Issues & 0.832*** & 0.222*** & 0.276*** \\
Genre: Opinion & 0.148** & 0.292*** & 0.076*** \\
Genre: Media & 0.230*** & -0.021*** & -0.089*** \\
Genre: International & 0.195*** & 0.080*** & 0.023** \\
Genre: Culture & 0.496*** & 0.118*** & -0.046*** \\
Genre: Lifestyle & 0.844*** & 0.270*** & 0.154*** \\
Genre: Panorama & 0.348*** & 0.183*** & 0.110*** \\
Genre: Podcast & 0.447** & 0.054** & -0.058 \\
Genre: Law & 0.241 & 0.242*** & 0.205*** \\
Genre: Sports & 0.546*** & 0.175*** & 0.114*** \\
Genre: Video & 0.671*** & 0.220*** & 0.124*** \\
Genre: Web & 0.625*** & 0.145*** & 0.085*** \\
Genre: Economy & 0.176*** & 0.027*** & 0.038*** \\
Genre: Science & 1.316*** & 0.192*** & 0.053*** \\ \hline
Dispersion & - & 0.744 & 0.220 \\
AIC & 55031 & 5850080 & 2883458 \\
BIC & 55335 & 5850452 & 2883830 \\
F1 & 0.0007 & - & - \\
RMSLE & - & 0.9590 & 0.7815 \\
\end{tabular}}
\label{tab:pinvote-regressions}
\end{table}

\begin{table}
\centering
\tbl{Calculated coefficients for Relative Voting Preference and the Comment Gap. ***: $p < 0.001$, **: $p < 0.01$, *: $p < 0.05$.}{
\begin{tabular}{l|cc}
Feature & \begin{tabular}[c]{@{}c@{}}Relative Voting\\Preference\end{tabular} & Comment Gap \\ \hline
Positive Sentiment & 0.044*** & 0.094*** \\
Negative Sentiment & -0.043*** & 0.007 \\
Lexical Diversity & -0.031*** & 0.579*** \\
Readability & -0.022*** & 0.156*** \\
Topical Similarity to Article & 0.042*** & -0.063*** \\
\# Punctuation Marks & 0.068*** & -0.116*** \\
\# Sentences & -0.051*** & 0.134*** \\
Text Uses 2nd Person Pronouns & -0.217*** & 0.434*** \\
Author Follower Count & 0.164*** & -0.062*** \\
Time Since Article Publication & 0.033*** & -1.224*** \\
Is Root Comment & -0.316*** & - \\
Comment Level in Tree & -0.240*** & - \\
Mean Upvotes in Discussion & 0.375*** & -0.174*** \\
Mean Downvotes in Discussion & -0.686*** & 0.755*** \\
\# Comments in Discussion & 0.043*** & -0.147*** \\
Genre: Women's Issues & -0.054* & 0.887*** \\
Genre: Opinion & 0.216*** & -0.068 \\
Genre: Media & 0.068*** & 0.162* \\
Genre: International & 0.057*** & 0.139** \\
Genre: Culture & 0.164*** & 0.332*** \\
Genre: Lifestyle & 0.117*** & 0.727*** \\
Genre: Panorama & 0.074*** & 0.274*** \\
Genre: Podcast & 0.113** & 0.334* \\
Genre: Law & 0.037 & 0.204 \\
Genre: Sports & 0.061*** & 0.485*** \\
Genre: Video & 0.095*** & 0.575*** \\
Genre: Web & 0.060*** & 0.565*** \\
Genre: Economy & -0.011 & 0.187*** \\
Genre: Science & 0.138*** & 1.178*** \\
\end{tabular}}
\label{tab:logodds-regression}
\end{table}

\subsubsection{Editors' Picks} \label{ii.-editors-picks}

We next turn to how different comment characteristics are associated with being being pinned as ``Editors' Picks'' (RQ1.2). For journalists' comment preferences, we consider the logistic regression coefficients in Table \ref{tab:pinvote-regressions}. An increase in a comment's Positive Sentiment is associated with increased odds of it being selected as an Editors' Pick, whereas an increase in Negative Sentiment is associated with (weakly significant) decreased odds of being selected. Journalists prefer more positive comments. Journalists also prefer more Lexically Diverse, higher Readability score comments with higher Number of Sentences, that use 2nd Person Pronouns from commenters with high Author Follower Count. Comments with high Number of Punctuation Marks are less likely to be selected as Editors' Picks (weakly significant), and there is no significant association between a comment's Topical Similarity to the article and whether it is selected as an Editors' Pick. For the article-level effects, firstly, the Number of Comments in Discussion is negatively associated with a comment being selected as an Editors' Pick---the number of picks tends to be limited, no matter how long the discussion. Both Mean Upvotes and Downvotes in Discussion are associated with increased odds of comments being selected as Editors' Picks---discussions with higher levels of engagement have more comments as Editors' Picks. On article genre, compared to the baseline category (Domestic); with the exception of Law (no significant difference), comments in articles of other genres have increased odds of being selected as an Editors' Pick. The strongest effects being for the Science, Lifestyle, and Women's Issues genres.

\subsubsection{The comment gap} \label{iii.-size-of-the-comment-gap}

\begin{figure*}
    \centering
    \includegraphics[width=\textwidth]{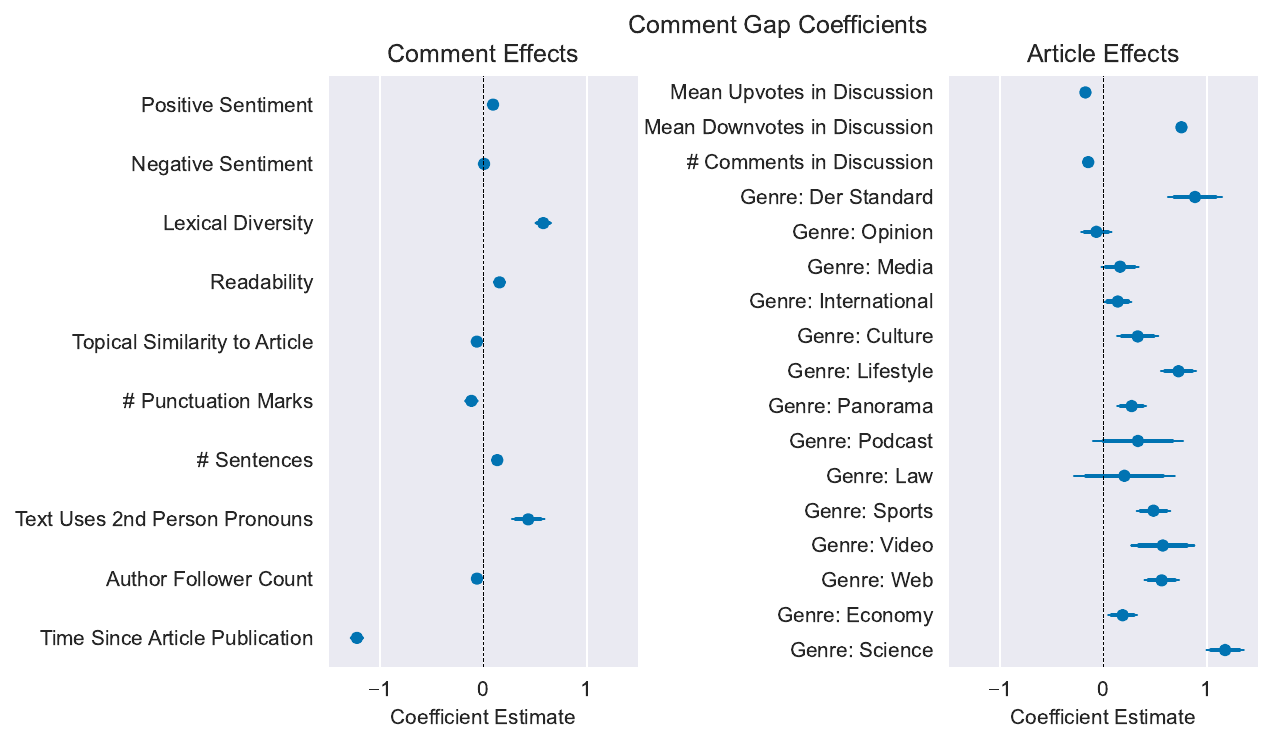}
    \caption{Regression coefficients  showing journalist preferences relative to readers for different comment characteristics. 95\% and 99\% confidence intervals indicated (many obscured by point size). Full data in Table \ref{tab:logodds-regression}.}
    \label{fig:commentgap-logodds}
\end{figure*}

Here we address RQ1.3, on the gap between journalist and reader preferences. As indicated previously, journalists and readers share several preferences in comments. Namely, for Positive Sentiment comments (and against Negative Sentiment comments) from commenters from high Author Follower Count, with several shared genre preferences. We also seek to assess the relative preferences of journalists vs readers, using the Comment Gap coefficients in Table \ref{tab:logodds-regression} and Figure \ref{fig:commentgap-logodds}. This compares the Editors' Picks odds against the Relative Voting Preferences previously calculated. Relative to readers, journalists have greater preference for Positive Comments and no significant difference in preference for Negative Comments. Editors have stronger preference for comments with higher Lexical Diversity, Readability score, Number of Sentences that use 2nd Person Pronouns. In contrast, relative to journalists, readers prefer comments with higher Topical Similarity to the Article, with more Punctuation Marks, from commenters with Higher Author Follower Count, made a longer Time Since Article Publication. For the article-level features, the negative score for Number of Comments in Discussion indicates that the number of Upvotes in a discussion scales more closely with discussion length compared to Downvotes and Editors' Picks. The negative(/positive) scores for Mean Upvotes(/Downvotes) trivially indicate that comments in discussions featuring more reader Upvotes(/Downvotes) are relatively preferred by readers(/journalists). Regarding article genre, compared to the baseline Domestic genre, the odds of a comment being selected as Editors' Pick increase compared to the Relative Voting Preference for most genres, with the exception of Opinion and Law. Journalists strongest relatively preferred comments are on Science, Lifestyle, and Video articles, whereas Readers strongest relatively preferred comments are on Opinion, Domestic, and International articles.

Additionally, in line with \citet{JuarezMiro2022}, we estimate the size of the news comment gap using the Jaccard similarity between top editor preferred comments and top reader preferred comments. The average Jaccard similarity between pinned comments and the top user-voted comments across all article discussions is 59.5\% for upvotes and 54.8\% for relative votes (upvotes - downvotes). This indicates a news comment gap of 40.5\% (or 45.2\%), in comparison to \citeauthor{JuarezMiro2022}'s finding of 82.8\%. In fact, since we have not accounted for the likely positive causal relationship between upvotes and Editors' Picks (pinned comments are more likely to be seen by readers and upvoted, and/or highly upvoted comments are more likely to be seen and picked by editors), this indicates the comment gap for \textit{Der Standard} is \textit{at least} 40.5\%.

\subsection{Ranking policies (RQ2)} \label{sorting-policies-rq2}

\begin{figure*}[h]
    \centering
    \includegraphics[width=\textwidth]{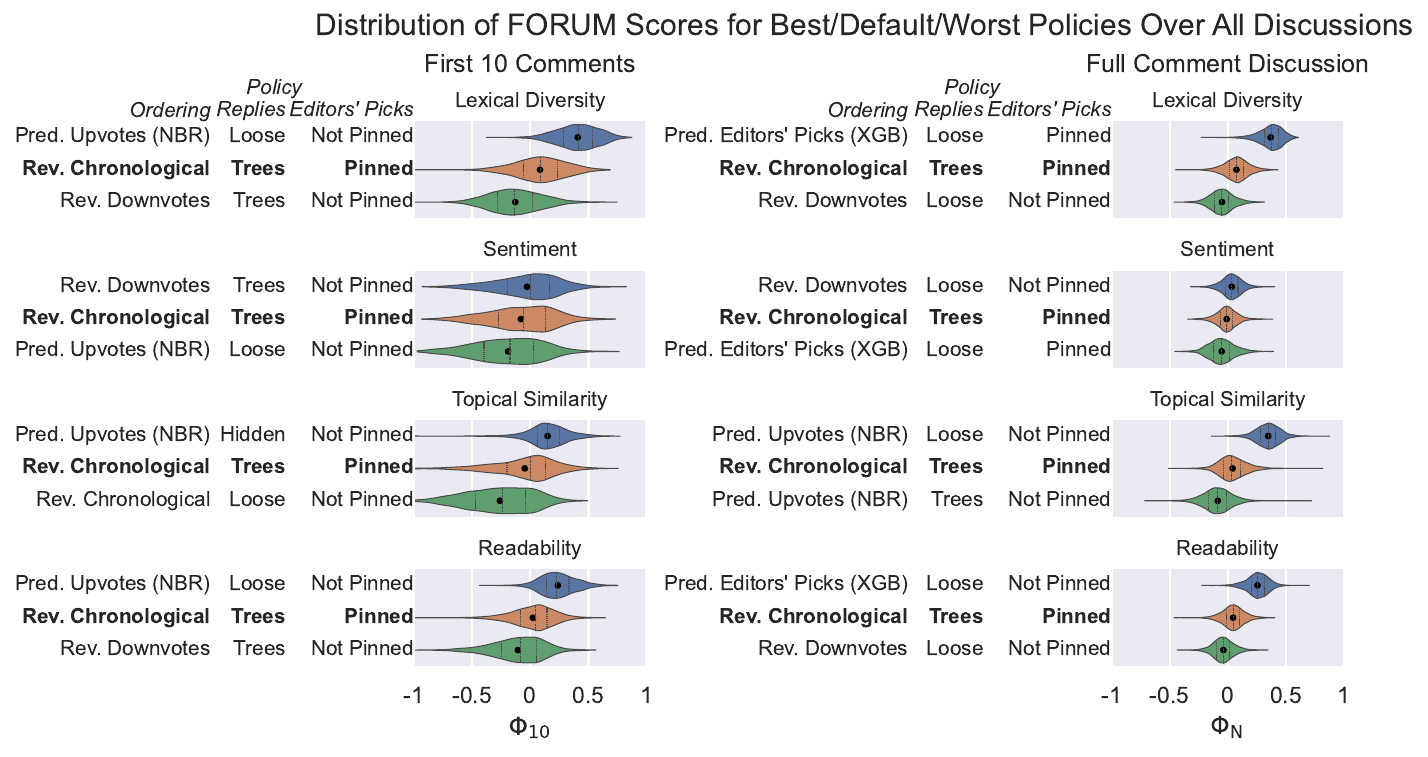}
    \caption{Distribution of FORUM score for best/default (reverse chronological, replies in trees, Editors' Picks pinned) / worst ranking policies in each feature over all discussions. Mean value (point) and quartiles (lines) indicated on each distribution.}
    \label{fig:qdistbdw}
\end{figure*}

\begin{figure*}[h]
    \centering
    \includegraphics[width=\textwidth]{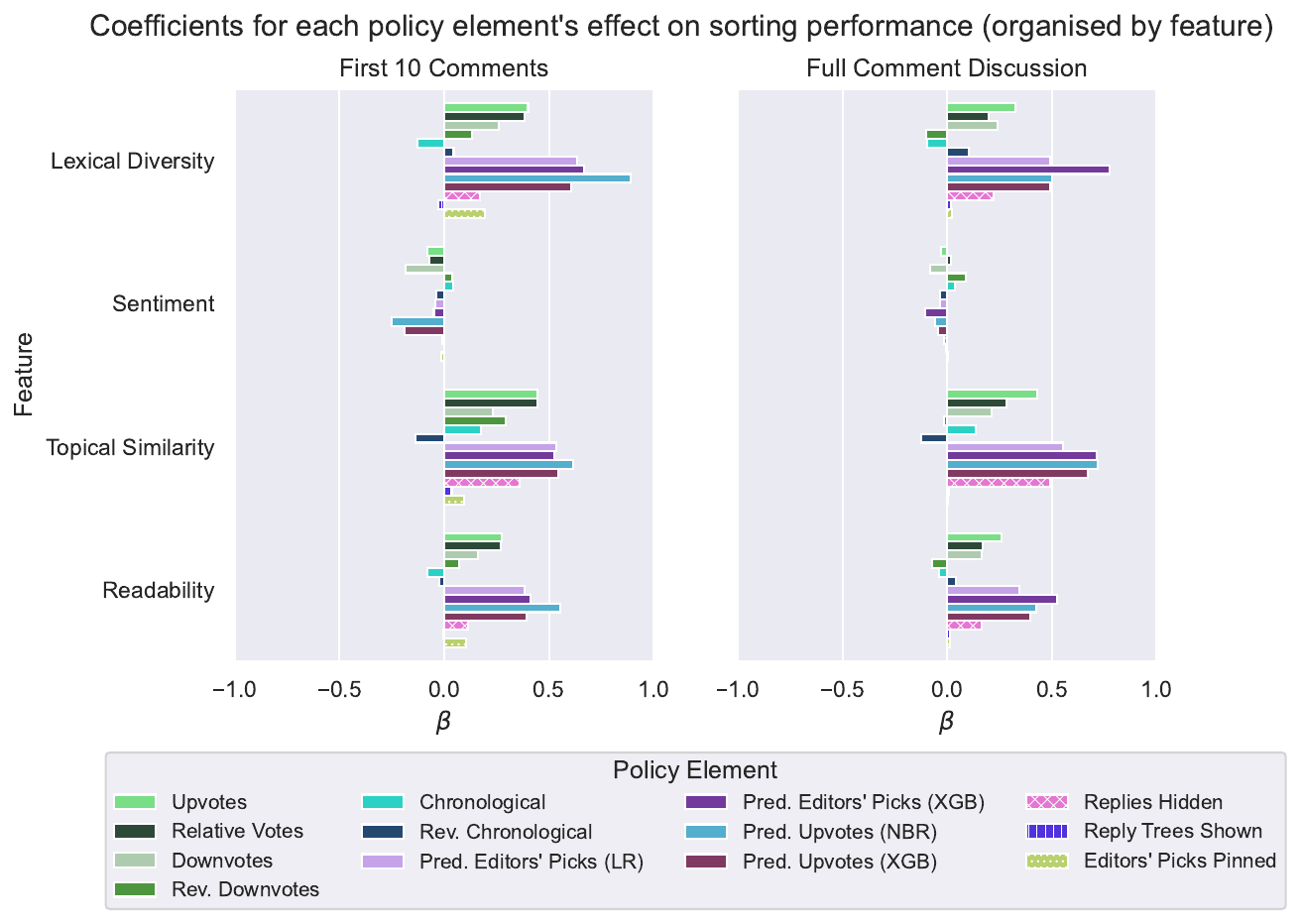}
    \caption{Main effect coefficients from beta regressions for the effect of implementing each policy element in a ranking algorithm on prioritising comments by each feature. Reference case is a random, replies loose, Editors' Picks unpinned policy. Bars are organised by outcome feature. Significance indicators omitted for visual clarity, but can be found in Appendix \ref{full-regressions}.}
    \label{fig:coef_feat}
\end{figure*}

We plot the distributions of FORUM for for all articles for the best, default, and worst ranking algorithms across the features of interest in Figure \ref{fig:qdistbdw}. Clearly, there can be large differences in the kinds of comments prioritised by different ranking algorithms. To understand how the different elements of a ranking algorithm (RQ2.1 primary ordering, RQ2.2 editor-based, RQ2.3 structure-based) contribute to how it (de-)prioritises certain kinds of comments, we interpret the coefficients from beta regressions presented in Figures \ref{fig:coef_feat}, \ref{fig:reply-int}, and \ref{fig:pin-int}. One can calculate the expected change on FORUM score from a coefficient with $\Delta \Phi = \frac{e^{\beta} - 1}{e^{\beta} + 1}$. Full regression tables are shown in Appendix \ref{full-regressions}.

\subsubsection{Primary ordering} \label{i.-user-based-sorting}

The 11 primary ordering categories are listed in Table \ref{tab:sortcats}. Considering lexical diversity, all the tested policies return more lexically diverse comments earlier than expected with random ordering, with the exception of Chronological over both the first 10 comments and the full discussion and Reverse Downvotes over the full discussion. Earlier comments are less lexically diverse than comments posted more recently. The least downvoted comments are more lexically diverse than other comments, but overall comments with more downvotes are more lexically diverse. The strongest of all effects is shown by implementing ordering by Predicted Upvotes (NBR) ($\beta_{10} = 0.890$) and Predicted Editors' Picks (XGB) ($\beta_N = 0.773$), which increase $\Phi$ from the baseline by 0.418, 0.368 respectively.

The picture for sentiment is slightly more mixed, with weaker effects in different directions. Most primary orderings across the first 10 comments and full discussion tend to slightly prioritise more negative sentiment comments than expected at random. However, Reverse Downvotes and Chronological prioritise more positive sentiment comments, and Relative Votes very slightly prioritises more positive comments over the full discussion. The strongest of all effects is shown by implementing ordering by Predicted Upvotes (NBR) ($\beta_{10} = -0.250$) and Predicted Editors' Picks (XGB) ($\beta_N = -0.107$), which decreases $\Phi$ from the baseline by 0.124, 0.053 respectively.

For topical similarity, all policies return more relevant comments earlier than expected with random ordering, with the exception of Reverse Chronological over both the first 10 comments and the full discussion and Reverse Downvotes (very weakly) over the full discussion. More recent comments are less topically similar to the article than comments posted earlier. The least downvoted comments are more topically similar to the article than other comments, but overall comments with fewer downvotes are less similar to the article. The strongest of all effects is shown by implementing ordering by Predicted Upvotes (NBR) ($\beta_{10} = 0.617$, $\beta_N = 0.719$), which increases $\Phi$ from the baseline by 0.299, 0.345 respectively.

Regarding readability, all policies return comments with higher SMOG reading age score earlier than expected with random ordering, with the exception of Chronological over both the first 10 comments and the full discussion, Reverse Chronological over the first 10 comments, and Reverse Downvotes over the full discussion. Both the earliest and most recent comments are of lower reading age score than those posted mid-discussion. Overall though, comments posted later are of higher reading age score. The least downvoted comments are of higher reading age score than others, but overall comments with more downvotes have a lower reading age score. The strongest effects are shown by implementing ordering by Predicted Upvotes (NBR)  over the first 10 comments ($\beta_{10} = 0.555$) and Predicted Editors' Picks (XGB) over the full discussion ($\beta_N = 0.522$), which increase $\Phi$ from the baseline by 0.271, 0.255 respectively.

\subsubsection{Editor-based ranking} \label{ii.-editor-based-sorting}

The editor-based sorting policy element controls whether comments selected by journalists as Editors' Picks are pinned to the top of the conversation, i.e. ranked ahead of all other comments. Pinning Editors' Picks to the top of a comment discussion has an insignificant to very weak effect on the sorted comments returned over the full discussion. This is unsurprising, given typically only 1-5 comments are pinned out of a discussion of several hundred comments. The effect is more notable, however, for lexical diversity, readability score, and topical similarity when only considering the first 10 displayed comments ($\beta_{10} = 0.192, 0.104, 0.092$). Pinning Editors' Picks then increases $\Phi_{10}$ above our baseline with no comment pinning by 0.096, 0.052, 0.046 respectively.

\subsubsection{Structure-based ranking} \label{iii.-structure-based-sorting}

Categories based on discussion structure determine how replies to comments are shown, if at all. Replies may be hidden, shown fixed in the reply tree structure, or independently ranked as loose comments (baseline).

For both the full discussion and first 10 comments, hiding replies increases the level of topical similarity, lexical diversity, and readability score seen earlier. Hiding replies also very weakly negatively affects prioritised sentiment over the full discussion. The strongest effects exist for hiding replies on topical similarity ($\beta_{10}=0.359$, $\beta_N=0.490$), equivalent to increasing $\Phi_{10}$, $\Phi_{N}$ above our baseline with loose comments by 0.178, 0.240 respectively. This aligns with expectations, since replies are one or more steps removed from responding to the news story itself, and are more likely to focus on addressing content in the comment they are responding to.

Showing full reply trees fixed in place typically has an insignificant to very weak effect on each of the features, when comparing to baseline. However, we soon see that it can make a substantial difference to FORUM score when considering the interaction effects with the different primary orderings.

\subsubsection{Interaction of sorting policy elements} \label{iv.-interaction}

\begin{figure}[h]
    \centering
    \includegraphics[width=\textwidth]{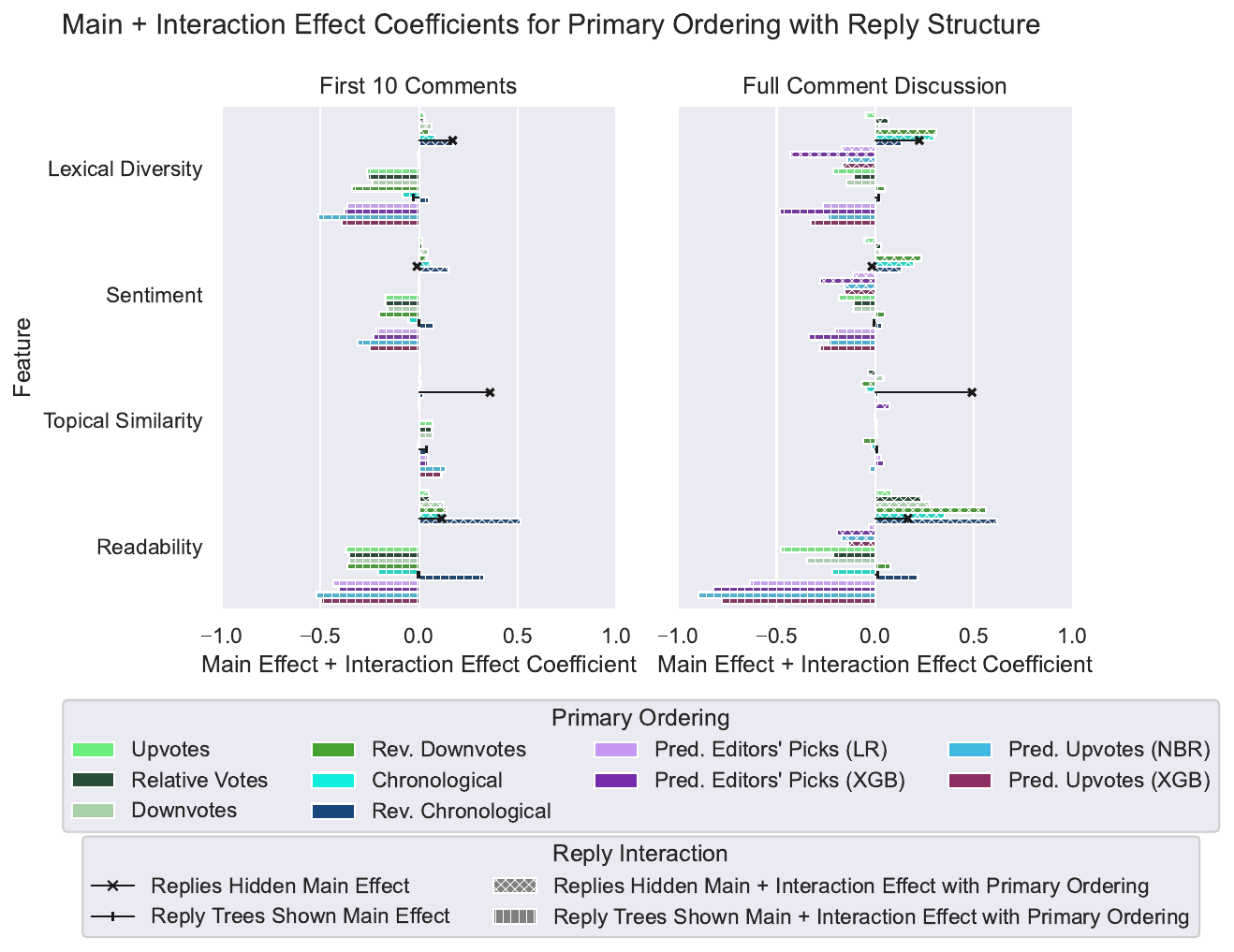}
    \caption{Main + interaction effect coefficients for primary ordering with reply structure. The lines show the main effect on a random primary ordering of hiding replies / fixing reply trees. The bars show the main + interaction effect on the outcome feature when hiding replies / fixing reply trees is applied to the alternative primary orderings. Reference case is a random, replies loose, Editors' Picks unpinned policy. Significance indicators omitted for visual clarity, but can be found in Appendix \ref{full-regressions}.}
    \label{fig:reply-int}
\end{figure}

As previously mentioned, the most parsimonious models include interaction terms between the 10 (non-baseline) primary ordering policies and total of 3 reply structure / pinned comment status variables. This results in 30 interaction terms in each regression. A number of these effects are comparable, if not larger, in size to the main effects previously presented, which means they may augment, weaken, nullify, or even reverse the previously observed main effects. A summary of these terms is presented in Table \ref{tab:reg-terms}, with their effects shown in Figures \ref{fig:reply-int} and \ref{fig:pin-int}. The full regression output, and more detailed discussion of these effects, is shown in Appendix \ref{full-regressions}. Ultimately, reply structure and pinning Editors' Picks has differential effects on the qualities of displayed comments, depending on the underlying primary comment ordering. The variability in magnitude and direction of these interaction effects illustrates the importance of considering every element of a particular ranking policy. The simplest way to compare average effects of specific policies is to consult Figure \ref{fig:qdistbdw}, and Figures \ref{fig:xsorting10}, and \ref{fig:xsortingN} from Appendix \ref{forum-distributions}.

\begin{figure}[h]
    \centering
    \includegraphics[width=\textwidth]{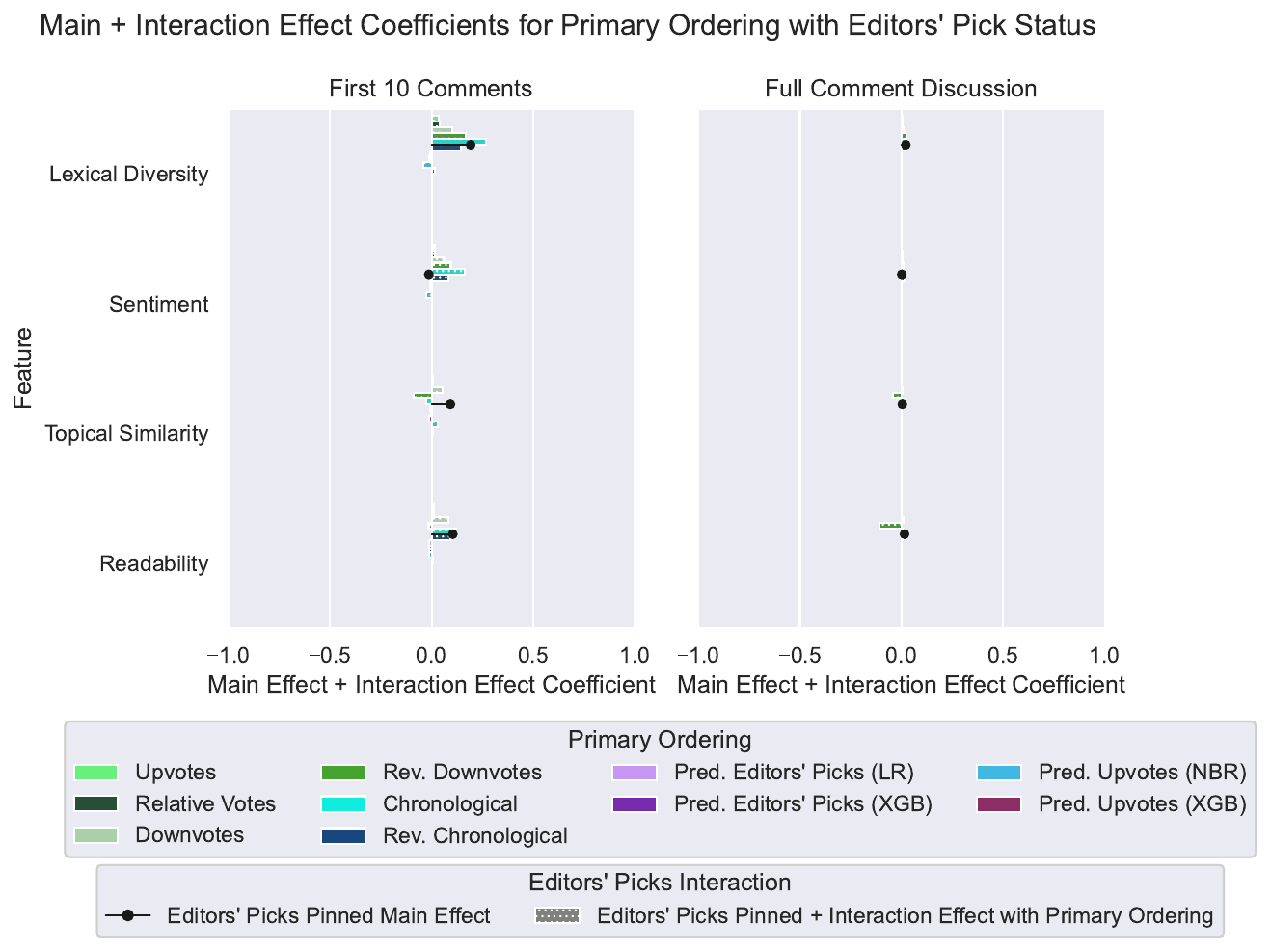}
    \caption{Main + interaction effect coefficients for primary ordering with Editors' Picks status. The lines show the main effect on a random primary ordering of pinning Editors' Picks. The bars show the main + interaction effect on the outcome feature when pinning Editors' Picks is applied to the alternative primary orderings. Reference case is a random, replies loose, Editors' Picks unpinned policy. Significance indicators omitted for visual clarity, but can be found in Appendix \ref{full-regressions}.}
    \label{fig:pin-int}
\end{figure}

\subsubsection{Notable other patterns} \label{v notable pattterns}

In considering all coefficients, there are some notable other patterns that come to light. Namely, differing policy performance for first 10 comments vs full discussion, typically strong effects for the predictive models, and a reverse ranking paradox.

It is interesting to note the difference in coefficients between the same policy elements over the first 10 presented comments vs the full discussion. Whilst they are fairly correlated, relative performance is not fully consistent, and occasionally inverts (e.g. Reverse Downvotes ordering for Lexical Diversity). This means the relationship between the target feature and feature(s) used for ordering is not linear, and cannot be adequately described by a single regular correlation measure. In the case of Reverse Downvotes and Lexical Diversity, this means the least downvoted comments are more lexically diverse than other comments, but overall, comments with fewer downvotes are less lexically diverse.

Many of the largest in magnitude coefficients are associated with the predictive models. Each of the models directly use the target features in predicting upvotes / pinned status, so perhaps it is not surprising to see the large coefficients as a result of the strong associations in the models between the target features and model objectives. This serves as a warning that the most extreme effects on the display of discussions occur as a result of predictive algorithms. Careful consideration of these effects must be made when predictive ranking algorithms are applied to online discussions.

It is perhaps counterintuitive that coefficients for reverse policies are not simply the inverse of those for the original ordering (e.g. Chronological vs Reverse Chronological). In the case of the first 10 comments, the amount the first 10 comments over/under-perform the random expected baseline feature score by one policy is not the same as the amount the first 10 comments of the reverse policy under/over-perform the baseline. Regarding the full discussion, a true reverse rank of comments in a discussion will produce an exact inverse FORUM score. The reasons for the deviation are that firstly the regression model specification, whilst most parsimonious, does not account for all interaction term variations. Secondly, for the vote-based orderings, comments with equal votes (typically 0) are  randomly placed. This means that Downvotes vs Reverse Downvote ranking policies applied to a discussion will not necessarily be the reverse of each other, so do not necessarily have exact inverted FORUM scores.

\section{Discussion} \label{discussion}

\subsection{The comment gap} \label{the-comment-gap}

Journalists and readers partly agree on which comments are valuable, both favouring positive comments from elite authors and dislike negative comments. However, readers also favour comments similar to the article, more (suspected informal use of) punctuation, posted later and disfavour comments with higher lexical diversity, readability score, number of sentences that directly address the author. Journalists also favour lexically diverse, high readability score, longer comments that directly address the author and disfavour comments with more punctuation made longer after article publication. Relative to each other, journalists prefer longer, more complex, earlier, direct responses, whereas readers prefer comments more similar to the article from elite authors. These factors drive the news comment gap.

Overall, there is an average similarity of 59.5\% between Editors' Pick comments and the most upvoted comments, and 54.8\% for relative votes. This is significantly higher than the 17.2\% overlap found by \citet{JuarezMiro2022} in the New York Times. While beyond the scope of this paper, and not possible to answer with our current dataset, the question about causality arises. do upvotes influence editors' selections, or do pinned and prominent comments attract more upvotes? Likely, both effects play a role, influenced by platform design. In \emph{Der Standard}'s forum, pinned comments appear first followed by the rest reverse-chronologically, unlike the NYT's solely reverse-chronological default. Most pinned comments were posted early (Appendix \ref{timedist}), likely gaining more votes due to their prominent position. Thus, the true comment gap may be at least 40.5\%.

On comment qualities, our findings partly align with \citeauthor{JuarezMiro2022}'s NYT study, which found readers favour confrontational comments, while we find readers dislike negative comments with 2\textsuperscript{nd} person pronouns. Both studies agree that journalists prefer positive comments---Editors' Picks in the NYT are ``more conciliatory [..], striving to promote agreement between commenters'' \cite[p 866]{JuarezMiro2022}. Additionally, both show journalists favour longer comments, whereas readers prefer shorter ones.

\subsection{Ranking is agenda-setting} \label{sorting-is-agenda-setting}

\textit{Der Standard} has three main levers for comment ranking: primary ordering, pinning comments, and showing replies. Primary ordering affects all comments (averaging 204 per discussion) and is highly impactful. However, since most users read only the first few comments, pinned comments hold substantial more standard curational agenda-setting power for editors. Additionally, two-thirds of comments are replies, and controlling them can systematically alter the discussion's structure. We developed the FORUM score to measure the impact of these choices.

\subsubsection{Primary orderings}

The primary orderings studied, including those used by \textit{Der Standard} and predictive models for upvotes / Editor's Picks, positively influence the display of lexically diverse, topically similar, and high readability score comments, but negatively affect prioritised comment sentiment. Various methods align with journalist and reader preferences, though their effectiveness varies. Default settings and the available options, while directionally similar, have differing impacts on discussion quality. Different ranking algorithms can contribute to `algorithmic agenda-setting', prioritising certain comment types and (mis)representing public reaction in a particular way. Journalists may choose to take advantage of this and align their comment ranking algorithms to prioritise comments that support their stories.

\subsubsection{Editor-based}

Pinning comments gives editors considerable agenda-setting power. \textit{Der Standard}'s forum shows pinned comments at the top, with an average of four per discussion, despite them being only 0.5\% of all comments. These pinned comments are disproportionately viewed. Our findings show that pinning Editors' Picks significantly impacts the FORUM score for the first 10 comments, similar to other ordering and reply structure policies. Pinning a few comments allows precise curation of discussions, unlike broader changes to comment ordering or reply displays. This more traditional editorial agenda-setting tool interacts with other algorithmic ranking policies. Further research could explore the effects of larger selections of Editors' Picks, i.e., more manually curated comment sections.

\subsubsection{Structure-based}

On average, only one-third of comments are root comments. We demonstrated how structure-based ranking affects comment order and discussion density (replies shown in trees vs. loose, or hidden). Hiding replies generally improves lexical diversity, topical similarity, and readability score of prioritised comments. However, interaction effects with primary orderings produce mixed results. Showing reply trees has weak or insignificant effects on random primary ordering but strong negative interaction effects with alternative orderings. The quality of replies depends on how root comments are prioritised, leading to varied interaction effects. Further research should explore how different reply structures affect conversation interpretability and reader experience, balancing high-quality comments with contextual discussions.

\subsection{Implications and future directions}

The default comment ranking on \textit{Der Standard} (reverse chronological, full reply trees, pinning Editors' Picks) performs close to the theoretical random baseline across studied features (see Fig. \ref{fig:qdistbdw}), indicating significant room to both improve, and worsen, the presentation of comments to users. We advise against solely optimising for one FORUM score; future work should focus on sophisticated, goal-oriented outcomes, considering ranking algorithms' effects on various features and, crucially, their as-yet unstudied effects on user behaviour. The strong effects from predictive ranking algorithms highlight both the potential, and dangers, of more sophisticated algorithmic comment ranking systems.

Our versatile FORUM score can be applied in future research, including different news platforms and experimental studies on ranking algorithms' impact on user behaviour. It can also be used to study ranking in various contexts where a finite set of items, each with measurable features, may be ranked and presented by some (potentially unknown) algorithm. For example, news articles, Reddit posts/comments, X/Twitter replies (with prioritised premium scubscribers), LinkedIn job postings, product listings on Amazon or other online marketplaces, etc. or the effects of different kinds of content moderation policies on social media.  

\subsection{Limitations} \label{limitations}

Comment ranking affects readers', authors', and voters' experiences. Since \textit{Der Standard}'s default sorting policy, which very few users change \citep{Lampe2014}, influences how comments are written and votes are cast. For this reason, there are certain limited causal inferences that can be made from some aspects of this work. Different ranking policies could result in different comments and user behaviours. Future research should explore how various ranking algorithms impact not just displayed comments but also user behaviour, (e.g., overall activity, confrontation, sentiment and tone of replies, etc.). This would likely require (natural) experiments on comment forums implementing different ranking algorithms.

Comments can be deleted by users or moderators, with only a deletion note left. This non-random deletion can introduce bias, as moderator-deleted comments are likely more controversial, and users may delete their comments if they receive backlash. However, only 2\% of comments in the dataset are deleted, so this bias is small.  A full year's data, rather than two months, would provide a more complete picture of commenting behaviour, accounting for seasonal news fluctuations. It is difficult to assess whether news coverage was unusual in those two months. However, both, genres and articles, do not seem to be overly focused on specific topics/events such as the 2015 refugee crisis or the 2020 COVID-19 pandemic. Moreover, one advantage of FORUM is that it is normalised to the available comments, so would be unaffected by these issues. Additional contextual data such as page views and editorial decisions would have enriched this research but were unavailable due to proprietary restrictions and interview requests were unfortunately declined by \textit{Der Standard}.

\section{Conclusion} \label{conclusion}

This paper quantitatively analysed the news comment gap between journalists and readers using 1.2 million comments. We examined the content valued by journalists versus readers in news comment forums and compared different ranking policies' impact on discussion presentation. Our work on German-language news complements previous English-focused research. Crucially, we also introduced the FORUM score to evaluate how well ranking algorithms prioritise comments based on specific features.

The results show that both readers and journalists value positive comments from elite authors and dislike negative comments. Journalists favour longer, more complex comments, made earlier that directly address the author and disfavour comments with more punctuation. Readers favour comments more similar to the article, more punctuation, made a longer time after article publication and disfavour more longer, complex comments that directly address the author. Compared to each other, journalists have stronger preference for positive, longer, more complex comments, posted earlier that directly address the author, whereas readers prefer comments that are similar to article content, with lots of (informal) punctuation, from elite authors.

Different ranking algorithms can dramatically alter the types of comments prioritised in a discussion, affecting features like lexical diversity, sentiment, topical similarity, and readability. Predictive ranking algorithms amplify these effects. Journalists exercise gatekeeping power through pinned Editors' Picks, retaining traditional editorial agenda-setting influence. Yet the algorithmic agenda-setting effects from other ranking algorithm elements are frequently even larger. While users align with some journalist preferences, catering to users' tastes in comment presentation more closely might be beneficial for engagement. Of course, selecting a single ``best'' sorting policy is likely impossible. Journalists must consider their values for comment sections and choose ranking algorithms accordingly. Researchers and readers should be able to critically evaluate discussions based on comment display. Different publications, news stories, and audiences may require varied ranking policies. To assist with these decisions, we have developed FORUM as a powerful, versatile method of evaluating the consequences of ranking algorithms, and it may be applied beyond the realm of news comments. Future UX research should also explore the trade-offs between displaying conversations in context and prioritising comment quality.

Measuring comment and discussion quality is crucial for fostering constructive democratic discourse, reducing information overload, moderating objectionable content, and representing public opinion accurately. Increasing legal scrutiny of UGC leads to higher risks for news organisations to host comment sections. In spite of this, news media should make the most of UGC in order to both fulfil their role as the fourth estate fostering democratic debate culture, as well as to increase engagement in a responsible way for economic gain. This paper's findings can inform design choices of news forums to achieve these objectives.

\newpage

\bibliographystyle{apacite}
\bibliography{bibliography}

\newpage

\appendix

\section{Supplementary ranking models} \label{rank-models}

For the purpose of implementing new ranking algorithms to test with FORUM, we build several models for predicted Editors' Picks and Upvotes using logistic regressions, negative binomial regressions and XGBoost models. Rather than model reader/journalist preferences, these models are designed to operate as `live' ranking algorithms that can take into account additional information from after the comment is created (full discussion structure, Editors' Pick status, Up/Downvotes).

\begin{figure}[h]
    \centering
    \includegraphics[]{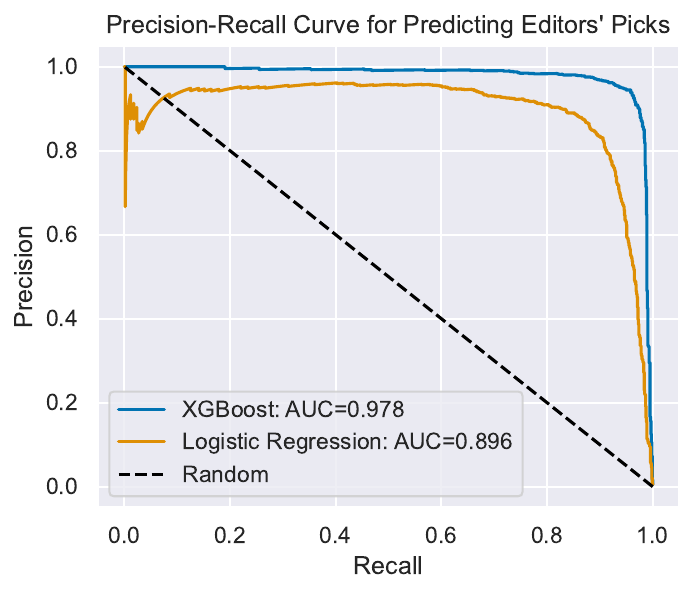}
    \caption{Precision-Recall curves for predicting Editors' Picks with different models.}
    \label{fig:pr-curve}
\end{figure}

Summaries of the logistic regression model for Editors' Picks and negative binomial regression model for Upvotes are shown in Table \ref{tab:pinvote-regressions-2}. We also train an XGBoost classifier for Editors' Picks and XGBoost regressor for Upvotes using the same features. All model performances are summarised in Tables \ref{tab:pin-performance} and \ref{tab:upvote-performance}, with the precision-recall curve for predicting Editors' Picks shown in Figure \ref{fig:pr-curve}. Root Mean Squared Log Error (RMSLE) is selected as performance measure and XGBoost loss function for most appropriate comparison to the negative binomial regression models.

\begin{table}
\centering
\tbl{Logistic and negative binomial regression model summaries with coefficients for a comment being selected as an Editors' Pick or receiving Upvotes. Note that coefficients in the Editors' Picks model for the structural factors are not fit here, since only root comments are eligible to be selected by journalists. ***: $p < 0.001$, **: $p < 0.01$, *: $p < 0.05$.}{
\begin{tabular}{l|cc}
Feature & Editors' Picks & Upvotes \\ \hline
Intercept & 26.624*** & 1.273*** \\
Positive Sentiment & 0.162*** & -0.019*** \\
Negative Sentiment & -0.076* & -0.009*** \\
Lexical Diversity & 0.367*** & 0.103*** \\
Readability & 0.042 & 0.038*** \\
Topical Similarity to Article & -0.029 & 0.006*** \\
\# Punctuation Marks & 0.105* & -0.045*** \\
\# Sentences & -0.000 & 0.012*** \\
Text Uses 2nd Person Pronouns & 0.108 & 0.035*** \\
Author Follower Count & 0.053 & 0.078*** \\
Time Since Article Publication & -0.182*** & -0.183*** \\
Is Root Comment & - & 0.020*** \\
Comment Level in Tree & - & -0.827*** \\
Mean Upvotes in Discussion & -0.520*** & 0.382*** \\
Mean Downvotes in Discussion & -0.352*** & 0.036*** \\
\# Comments in Discussion & -0.715*** & -0.016*** \\
Genre: Women's Issues & -0.034 & 0.156*** \\
Genre: Opinion & 0.006 & 0.252*** \\
Genre: Media & 0.160 & -0.028*** \\
Genre: International & 0.073 & 0.038*** \\
Genre: Culture & 0.400* & 0.101*** \\
Genre: Lifestyle & 0.488*** & 0.206*** \\
Genre: Panorama & 0.034 & 0.124*** \\
Genre: Podcast & 0.760** & 0.039* \\
Genre: Law & 0.049 & 0.161*** \\
Genre: Sports & 0.329* & 0.127*** \\
Genre: Video & 0.084 & 0.152*** \\
Genre: Web & 0.396** & 0.072*** \\
Genre: Economy & 0.033 & 0.009* \\
Genre: Science & 1.213*** & 0.114*** \\
Comment Pinned & - & 1.942*** \\
Is Leaf Comment & -8.191*** & -0.311*** \\
Size of Comment Tree & 28.711*** & 0.263*** \\
Height of Comment Tree & -1.427*** & -0.041*** \\
\# Replies to Comment & -13.993*** & -0.025*** \\
Comment Downvotes & 0.462*** & 0.042*** \\
Comment Upvotes & 2.672*** & - \\ \hline
Dispersion & - & 0.220 \\
AIC & 10875 & 2883458 \\
BIC & 11243 & 2883830 \\
F1 & 0.8645 & - \\
RMSLE & - & 0.9100 \\
\end{tabular}}
\label{tab:pinvote-regressions-2}
\end{table}

\begin{table}[h]
\centering
\tbl{Performance for the different Editors' Picks models. Model 1 is taken from Section \ref{the-comment-gap-rq1}, whereas models 2 and 3 represent the predictive ranking algorithms detailed here and used in Section \ref{sorting-policies-rq2}.}{
\begin{tabular}{l|lll}
Model       & 1      & 2     & 3     \\ \hline
Algorithm & \begin{tabular}[c]{@{}l@{}}Logistic\\ Regression\end{tabular} & \begin{tabular}[c]{@{}l@{}}Logistic\\ Regression\end{tabular} & XGBoost \\
\# Features & 27     & 33    & 33    \\ \hline
Accuracy    & 0.984  & 0.996 & 0.999 \\
Precision   & 1.000  & 0.912 & 0.949 \\
Recall      & 0.0003 & 0.821 & 0.944 \\
F1          & 0.0007 & 0.865 & 0.946
\end{tabular}}
\label{tab:pin-performance}
\end{table}

\begin{table}[h]
\centering
\tbl{Performance for the different Upvotes models. Model 1 is taken from Section \ref{the-comment-gap-rq1}, whereas models 2 and 3 represent the predictive ranking algorithms detailed here and used in Section \ref{sorting-policies-rq2}.}{
\begin{tabular}{l|lll}
Model       & 1     & 2     & 3     \\ \hline
Algorithm & \begin{tabular}[c]{@{}l@{}}Neg. Binomial\\ Regression\end{tabular} & \begin{tabular}[c]{@{}l@{}}Neg. Binomial\\ Regression\end{tabular} & XGBoost \\
\# Features & 29    & 35    & 35    \\ \hline
RMSLE       & 0.959 & 0.910 & 0.871
\end{tabular}}
\label{tab:upvote-performance}
\end{table}

\section{Full FORUM Distributions}
\label{forum-distributions}
\begin{landscape}
\begin{figure}
    \centering
    \includegraphics[width=\linewidth]{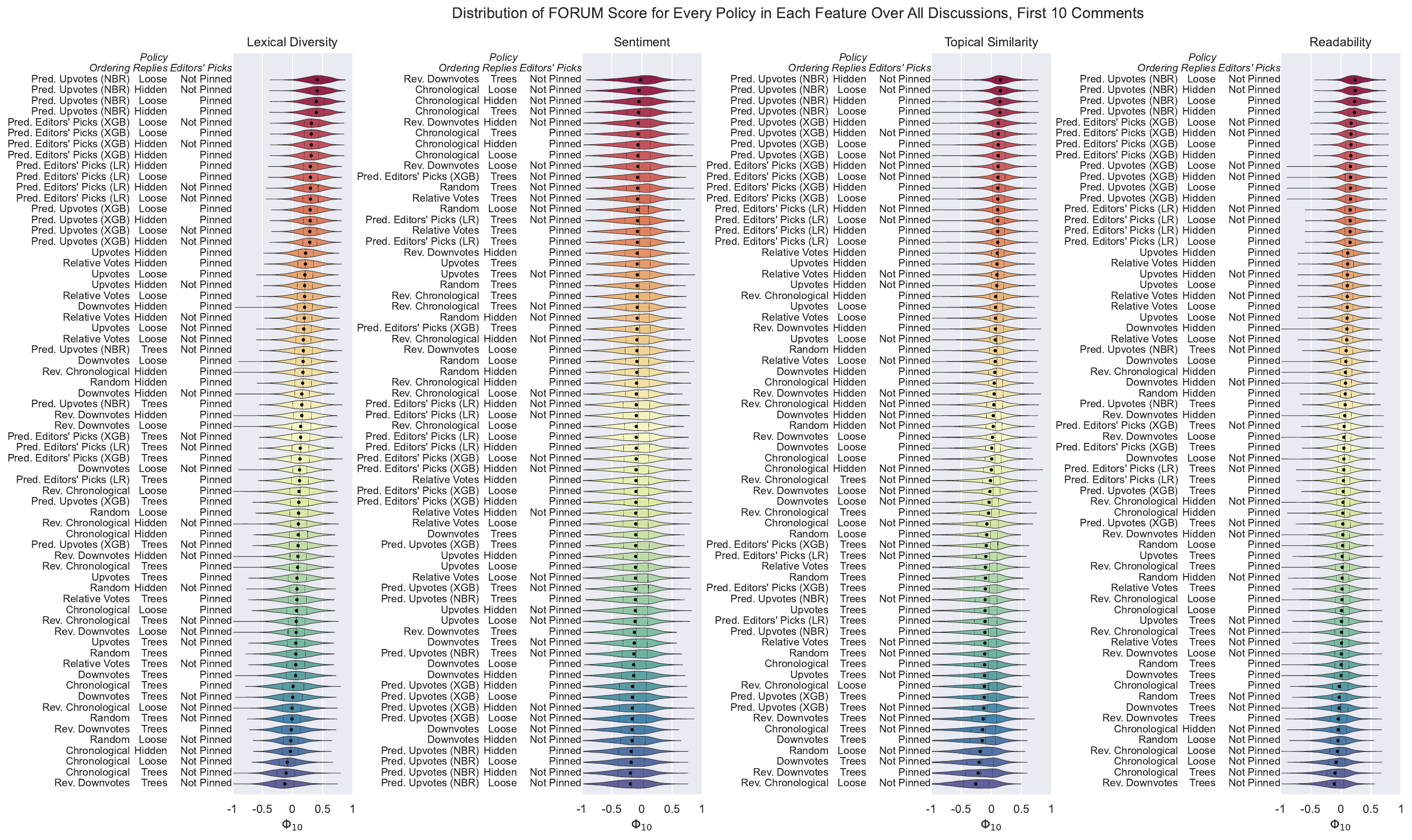}
    \caption{Distribution of FORUM score for every ranking policy in each feature over the first 10 comments from all discussions. Default policy (reverse chronological, replies in trees, Editors' Picks pinned) shown in bold. Mean value (point) and quartiles (lines) indicated on each distribution.}
    \label{fig:xsorting10}
\end{figure}
\end{landscape}

\begin{landscape}
\begin{figure}
    \centering
    \includegraphics[width=\linewidth]{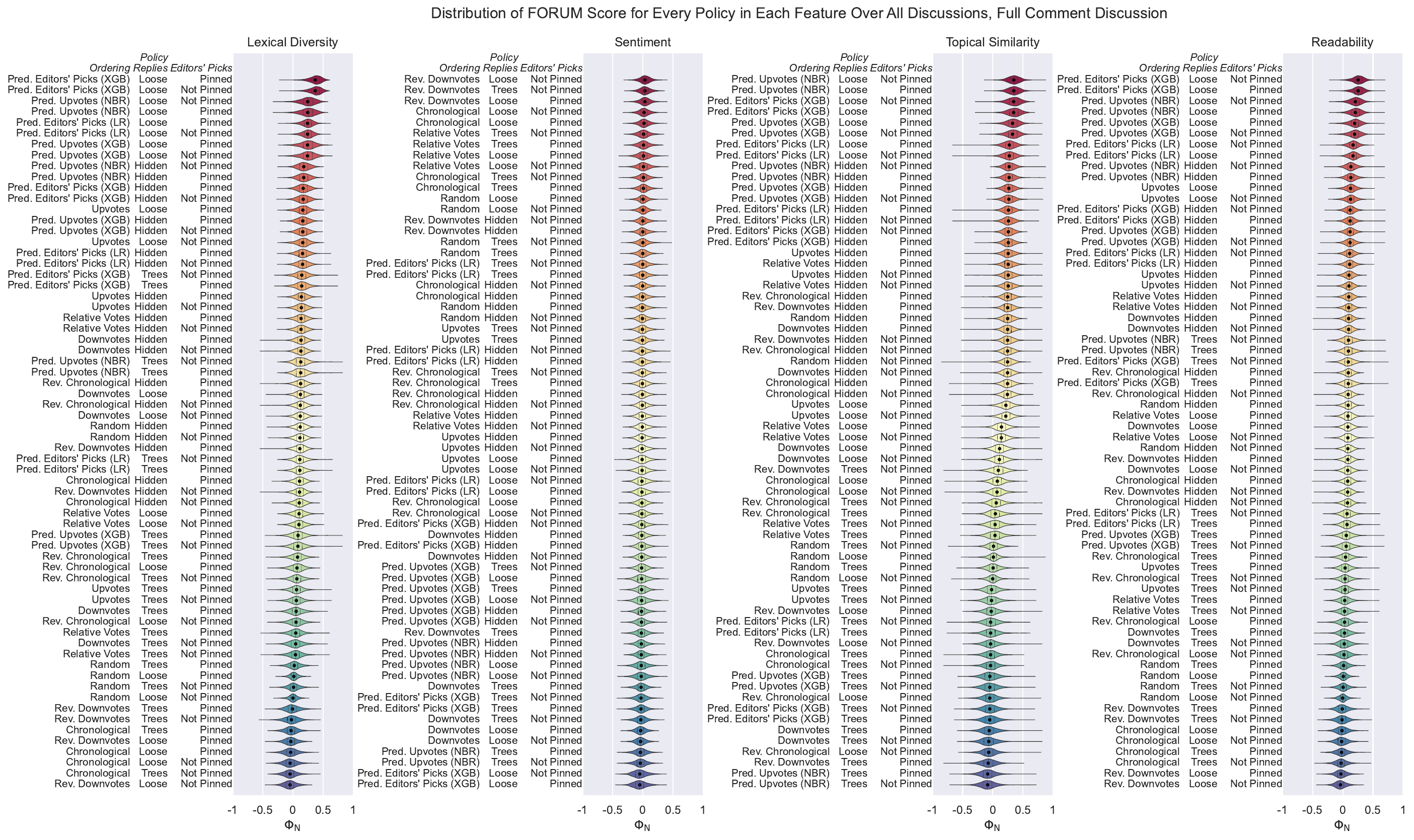}
    \caption{Distribution of FORUM score for every ranking policy in each feature from all discussions evaluated over full discussion. Default policy (reverse chronological, replies in trees, Editors' Picks pinned) shown in bold. Mean value (point) and quartiles (lines) indicated on each distribution.}
    \label{fig:xsortingN}
\end{figure}
\end{landscape}

We show the distribution of FORUM scores for all available ranking policies in Figures \ref{fig:xsorting10} and \ref{fig:xsortingN}.

\section{Beta Regression Model Selection}
\label{regression-specifications}
\begin{figure*}[h]
    \centering
    \includegraphics[width=\textwidth]{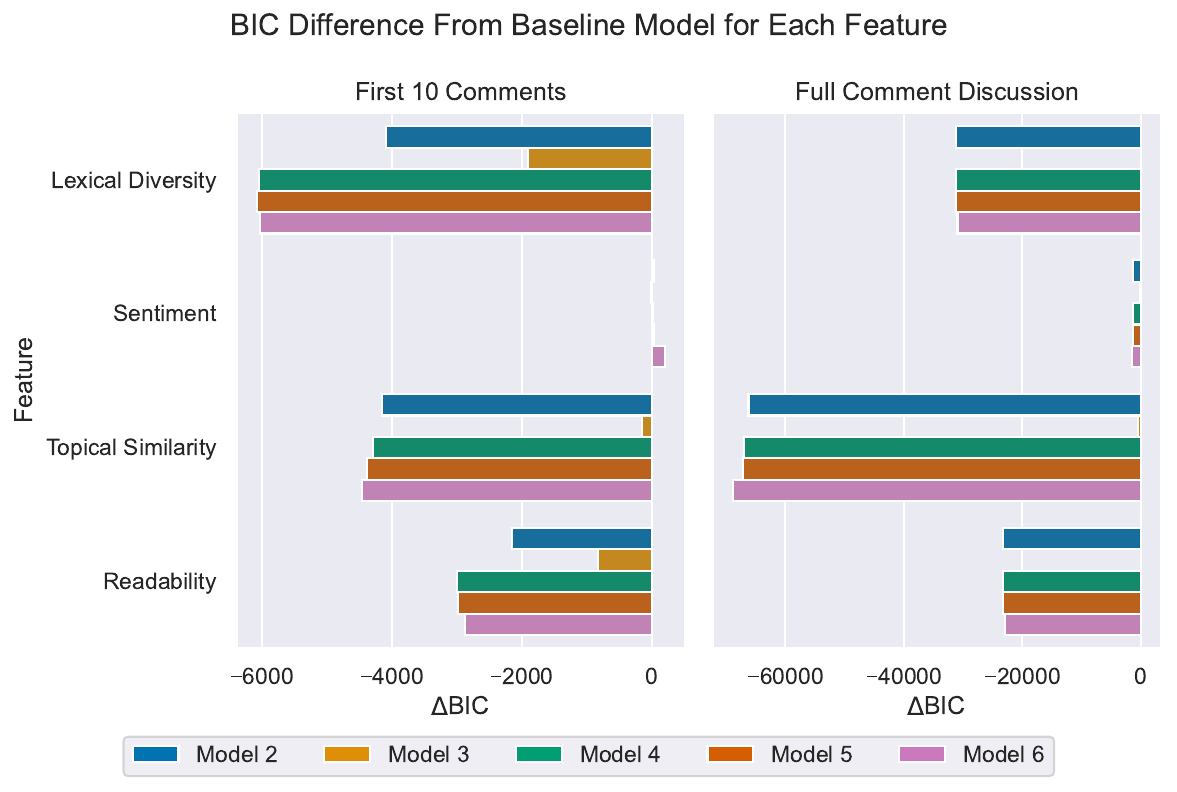}
    \caption{Difference between BIC scores for each model compared to baseline model 1.}
    \label{fig:BICs}
\end{figure*}

To model how FORUM score depends on each ranking policy element, we considered a number of beta regression specifications. The six tested model specifications, across four features and two discussion lengths (48 total models), and the main/interaction effect terms are summarised in Table \ref{tab:model-specs}. We evaluate the Bayesian information criterion (BIC) to determine the most parsimonious model(s). The BIC scores are shown in Table \ref{tab:BICs}, and their performance relative to baseline model 1 (no interaction effects) shown in Figure \ref{fig:BICs}. Unfortunately, there is no single best model specification across all features and discussion lengths. For the purposes of effective communication of results, we choose a single model to proceed with---Model 4. This model is best in one output, 2\textsuperscript{nd} best in three outputs, and 3\textsuperscript{rd} best in the remaining four. For each output it is still close in performance to the best model. The next best performing specification across all outputs---Model 6---is best in 3/8 outputs and 3\textsuperscript{rd} best in two outputs, but 4\textsuperscript{th} best in another two, and worst in one, where it is some way off the best model.

\begin{table}[h]
\tbl{Regression model compositions that were trialled. All considered models contain main effect terms for primary ordering, reply structure, and Editors' Picks status, but different interaction terms.}{
\begin{tabular}{l|cccccc}
                                                                                                                              & \multicolumn{6}{c}{Model}                                             \\
Policy Element Terms                                                                                                          & 1         & 2         & 3         & 4         & 5         & 6         \\ \hline
Primary Ordering                                                                                                              & \ding{52} & \ding{52} & \ding{52} & \ding{52} & \ding{52} & \ding{52} \\
Reply Structure                                                                                                               & \ding{52} & \ding{52} & \ding{52} & \ding{52} & \ding{52} & \ding{52} \\
Editors' Picks Status                                                                                                         & \ding{52} & \ding{52} & \ding{52} & \ding{52} & \ding{52} & \ding{52} \\ \hline
\begin{tabular}[c]{@{}p{3cm}@{}} Primary Ordering $\times$ \\ Reply Structure  \end{tabular}                                  & \ding{56} & \ding{52} & \ding{56} & \ding{52} & \ding{52} & \ding{52} \\ \hline
\begin{tabular}[c]{@{}p{3cm}@{}} Primary Ordering $\times$ \\ Editors' Picks Status    \end{tabular}                          & \ding{56} & \ding{56} & \ding{52} & \ding{52} & \ding{52} & \ding{52} \\ \hline
\begin{tabular}[c]{@{}p{3cm}@{}} Reply Structure $\times$ \\ Editors' Picks Status   \end{tabular}                            & \ding{56} & \ding{56} & \ding{56} & \ding{56} & \ding{52} & \ding{52} \\ \hline
\begin{tabular}[c]{@{}p{3cm}@{}} Primary Ordering $\times$ \\ Reply Structure $\times$ \\ Editors' Picks Status \end{tabular} & \ding{56} & \ding{56} & \ding{56} & \ding{56} & \ding{56} & \ding{52}
\end{tabular}}
\label{tab:model-specs}
\end{table}

\begin{table*}[h]
\centering
\tbl{BIC scores for beta regression fit for each model on each feature and number of comments considered. Best models indicated in bold, second best indicated in italics.}{
\begin{tabular}{l|c|rrrrrr}
 & & \multicolumn{6}{c}{Model} \\ 
Feature & n & 1 & 2 & 3 & 4 & 5 & 6 \\ \hline
Lexical Diversity & 10 & -288796 & -292887 & -290700 & \textit{-294836} & \textbf{-294867} & -294822 \\
Sentiment & 10 & \textit{-170377} & -170345 & \textbf{-170390} & -170358 & -170339 & -170166 \\
Topical Similarity & 10 & -148677 & -152821 & -148822 & -152963 & \textit{-153059} & \textbf{-153135} \\
Readability & 10 & -321775 & -323927 & -322600 & \textbf{-324763} & \textit{-324758} & -324646 \\ \hline
Lexical Diversity & N & -507087 & \textbf{-538298} & -507043 & \textit{-538269} & -538260 & -538039 \\
Sentiment & N & -575856 & -577116 & -575925 & -577187 & \textit{-577198} & \textbf{-577299} \\
Topical Similarity & N & -442190 & -508419 & -442621 & -509120 & \textit{-509382} & \textbf{-511008} \\
Readability & N & -558329 & \textbf{-581645} & -558244 & \textit{-581565} & -581547 & -581315 \\
\end{tabular}}
\label{tab:BICs}
\end{table*}

\begin{table}[h]
\centering
\tbl{BIC ranks for beta regression fit for each model on each feature and number of comments considered.}{
\begin{tabular}{l|c|rrrrrr}
 & & \multicolumn{6}{c}{Model} \\ 
Feature & n & 1 & 2 & 3 & 4 & 5 & 6 \\ \hline
Lexical Diversity & 10 & 6 & 4 & 5 & 2 & 1 & 3 \\
Sentiment & 10 & 2 & 4 & 1 & 3 & 5 & 6 \\
Topical Similarity & 10 & 6 & 4 & 5 & 3 & 2 & 1 \\
Readability & 10 & 6 & 4 & 5 & 1 & 2 & 3 \\ \hline
Lexical Diversity & N & 5 & 1 & 6 & 2 & 3 & 4 \\
Sentiment & N & 6 & 4 & 5 & 3 & 2 & 1 \\
Topical Similarity & N & 6 & 4 & 5 & 3 & 2 & 1 \\
Readability & N & 5 & 1 & 6 & 2 & 3 & 4 \\
\end{tabular}}
\label{tab:BICranks}
\end{table}

\section{Full Beta Regression Model Output}
\label{full-regressions}
We discuss the interaction effects for the FORUM score beta regressions visualised in Figure \ref{fig:reply-int}. The main effect of hiding replies on a random ordering baseline policy typically had a positive effect on displayed comments outcome features, ($ -0.010 \le \beta_{10} \le 0.359$, $-0.017 \le \beta_{N} \le 0.490$), with the exception of sentiment for which there was no effect or a very weak negative effect. With alternative primary orderings, we tend to see that the effect of hiding replies on lexical diversity is less positive, or even reversed so as to become negative, compared to that from random ordering. Notably, hiding replies when using of the predictive algorithm primary orderings has a substantial negative effect on displayed lexical diversity, sentiment, and readability score over the full discussion. For sentiment, the main effect term was $\approx 0$, but the interaction terms with alternative orderings show a wider range of positive and negative (with the predictive algorithms on the full discussion) effects. Hiding replies had a positive effect on topical similarity for random primary ordering, however this is broadly nullified across many other orderings. Hiding replies had a weakly positive effect on readability score with random primary ordering, but the effects vary with alternative primary orderings. Over the first 10 comments it has a markedly more positive effect on readability score when applied to reverse chronological ordering, and a slightly weakened / null effect when using the other orderings. The effects are more extreme when considering the full comment discussion, both positively and negatively.

Considering Figure \ref{fig:reply-int} again, the main effect of showing reply trees was typically very small, if not insignificant, for all outcome features when applied to a random ordering baseline policy ($-0.027 \le \beta_{10} \le 0.038$, $-0.006 \le \beta_{N} \le 0.018$). Yet with alternative primary orderings, showing full reply trees can have substantial positive and negative effects on the prioritisation of comments with certain features. For a random ordering we observed a null effect on lexical diversity for showing replies in trees. However, for almost all other primary orderings we find that showing replies in trees has moderate strong negative effect on lexical diversity. Similarly, the null effect on sentiment of showing replies in trees for random ordering is replaced by moderate negative effects for almost all other primary orderings. The null to weak positive effect on topical similarity for random ordering of showing full reply trees is replaced by weak positive effects with alternative orderings over the first 10 comments, and mixed null to weak effects over the full discussion. The null effect of showing reply trees on readability score when applied to random ordering is replaced by typically moderate to strong negative effects when using the other primary orderings. Exceptions to this are the positive effects when applied to reverse downvotes over the full discussion and reverse chronological over both the first 10 comments and full discussion.

For pinning Editors' Picks (Figure \ref{fig:pin-int}), the main effect over the full comment discussion was insignificant to very weak ($1\times 10^{-4} \le \beta_{N} \le 0.021$) and, with a few exceptions, this was reflected in the interaction terms ($-0.111 \le \beta_{N} \le 0.004$). When displaying the first 10 comments, pinning Editors' Picks had a positive effect on readability score and lexical diversity ($\beta_{10} = 0.015, 0.021$) when applied to the random ordering baseline policy. With alternative primary orderings, we see negative interaction terms that tend to slightly weaken or nullify the main effect term. In these cases, pinning Editors' Picks has slightly less of an effect on prioritised comment features when applied to (non-random) alternative primary orderings. Notably, for chronological primary ordering, pinning Editors' Picks increases the lexical diversity scores amongst the first 10 comments more than what we see with random primary ordering. We previously observed a null effect of pinning Editors' Picks on sentiment, but we now observe weak positive effects for the non-predictive models over the first 10 comments.

\begin{figure*}[h]
    \centering
    \includegraphics[width=\textwidth]{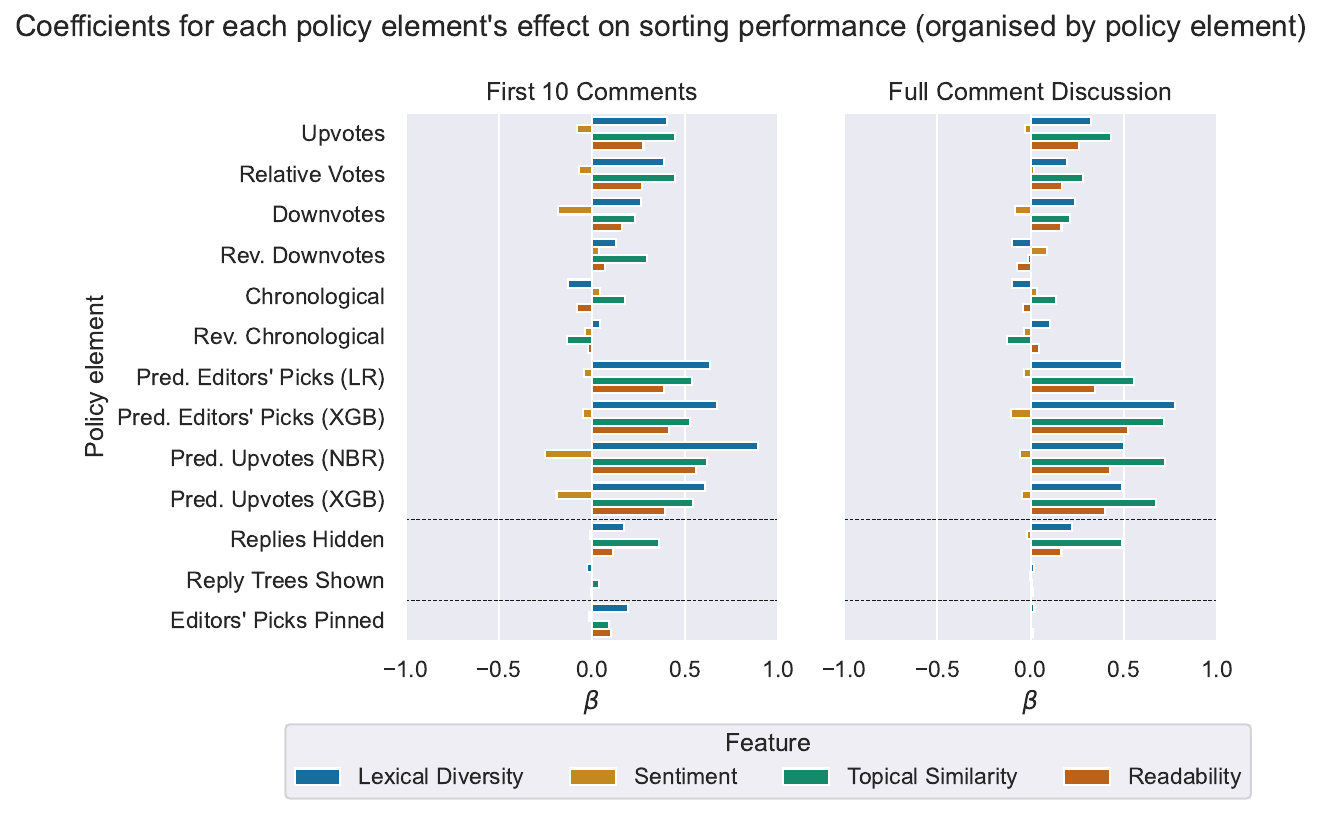}
    \caption{Main effect coefficients from beta regressions for the effect of implementing each policy element in a ranking algorithm on prioritising comments by each feature. Reference case is a random, replies loose, Editors' Picks unpinned policy. Bars are organised by policy element and are the same data as in Figure \ref{fig:coef_feat}.  Significance indicators omitted for visual clarity, but can be found in Appendix \ref{full-regressions}.}
    \label{fig:coef_pol}
\end{figure*}

Here, we also present the full regression output from the 8 main beta regressions studied in Section \ref{sorting-policies-rq2}. Results are shown in Tables \ref{tab:results-lexdiv_10}-\ref{tab:results-smog_N}. We also show an adapted version of Figure \ref{fig:coef_feat} in Figure \ref{fig:coef_pol}, where bars indicating regression coefficients are organised by policy element, rather than outcome feature.

\begin{table}[]
\centering
\tbl{Regression output for modelling FORUM score for Lexical Diversity over the first 10 comments (***: $p<0.001$, **: $p<0.01$, *: $p<0.05$, .: $p \ge 0.05$).}{
\begin{tabular}{c|c|c|c|c|c}
Outcome & Sample Size & AIC & BIC & Log-Likelihood & Dispersion ($\phi$) \\ \hline 
$\text{FORUM}^{\text{Lexical Diversity}}_{\text{10}}$ & 172,590 & -295289.0 & -294836.4 & 147689.5 & 21.6 \\
\end{tabular}}

\ 

\ 

\resizebox{\textwidth}{!}{
\begin{tabular}{l|r|r|r|c}
Variable & Estimate & Std. Error & $z$-value & $p$-value \\
\hline
---\textit{Main Effects} & & & & \\
(Intercept) & -0.0286517 & 0.0067262 & -4.26 & $2.05\times 10 ^ {-05}$*** \\
Downvotes & 0.2621861 & 0.0095443 & 27.47 & $< 2\times 10 ^ {-16}$*** \\
Rev. Downvotes & 0.1308048 & 0.0095217 & 13.74 & $< 2\times 10 ^ {-16}$*** \\
Upvotes & 0.3997788 & 0.0095808 & 41.73 & $< 2\times 10 ^ {-16}$*** \\
Relative Votes & 0.3852549 & 0.0095761 & 40.23 & $< 2\times 10 ^ {-16}$*** \\
Rev. Chronological & 0.0450165 & 0.0095139 & 4.73 & $2.23\times 10 ^ {-06}$*** \\
Chronological & -0.1280750 & 0.0095172 & -13.46 & $< 2\times 10 ^ {-16}$*** \\
Pred. Editors' Picks (LR) & 0.6342036 & 0.0096931 & 65.43 & $< 2\times 10 ^ {-16}$*** \\
Pred. Editors' Picks (XGB) & 0.6680616 & 0.0097136 & 68.78 & $< 2\times 10 ^ {-16}$*** \\
Pred. Upvotes (NBR) & 0.8900699 & 0.0098781 & 90.11 & $< 2\times 10 ^ {-16}$*** \\
Pred. Upvotes (XGB) & 0.6049027 & 0.0096757 & 62.52 & $< 2\times 10 ^ {-16}$*** \\
Replies Hidden & 0.1705937 & 0.0082637 & 20.64 & $< 2\times 10 ^ {-16}$*** \\
Reply Trees Shown & -0.0268341 & 0.0082413 & -3.26 & $0.00113$** \\
Editors' Picks Pinned & 0.1920458 & 0.0067426 & 28.48 & $< 2\times 10 ^ {-16}$*** \\
---\textit{Reply Structure Interactions} & & & & \\
Downvotes : Replies Hidden & -0.1076317 & 0.0117310 & -9.17 & $< 2\times 10 ^ {-16}$*** \\
Rev. Downvotes : Replies Hidden & -0.1229038 & 0.0116979 & -10.51 & $< 2\times 10 ^ {-16}$*** \\
Upvotes : Replies Hidden & -0.1443209 & 0.0117700 & -12.26 & $< 2\times 10 ^ {-16}$*** \\
Relative Votes : Replies Hidden & -0.1423499 & 0.0117643 & -12.10 & $< 2\times 10 ^ {-16}$*** \\
Rev. Chronological : Replies Hidden & 0.0008734 & 0.0116884 & 0.07 & $0.9404$. \\
Chronological : Replies Hidden & -0.0899816 & 0.0116741 & -7.71 & $1.28\times 10 ^ {-14}$*** \\
Pred. Editors' Picks (LR) : Replies Hidden & -0.1705955 & 0.0119029 & -14.33 & $< 2\times 10 ^ {-16}$*** \\
Pred. Editors' Picks (XGB) : Replies Hidden & -0.1778937 & 0.0119271 & -14.92 & $< 2\times 10 ^ {-16}$*** \\
Pred. Upvotes (NBR) : Replies Hidden & -0.1713304 & 0.0121369 & -14.12 & $< 2\times 10 ^ {-16}$*** \\
Pred. Upvotes (XGB) : Replies Hidden & -0.1733853 & 0.0118883 & -14.58 & $< 2\times 10 ^ {-16}$*** \\
Downvotes : Reply Trees Shown & -0.2079061 & 0.0116762 & -17.81 & $< 2\times 10 ^ {-16}$*** \\
Rev. Downvotes : Reply Trees Shown & -0.3121888 & 0.0116696 & -26.75 & $< 2\times 10 ^ {-16}$*** \\
Upvotes : Reply Trees Shown & -0.2378711 & 0.0117024 & -20.33 & $< 2\times 10 ^ {-16}$*** \\
Relative Votes : Reply Trees Shown & -0.2298533 & 0.0116992 & -19.65 & $< 2\times 10 ^ {-16}$*** \\
Rev. Chronological : Reply Trees Shown & 0.0754212 & 0.0116594 & 6.47 & $9.88\times 10 ^ {-11}$*** \\
Chronological : Reply Trees Shown & -0.0547763 & 0.0116621 & -4.70 & $2.64\times 10 ^ {-06}$*** \\
Pred. Editors' Picks (LR) : Reply Trees Shown & -0.3346886 & 0.0117866 & -28.40 & $< 2\times 10 ^ {-16}$*** \\
Pred. Editors' Picks (XGB) : Reply Trees Shown & -0.3520601 & 0.0118030 & -29.83 & $< 2\times 10 ^ {-16}$*** \\
Pred. Upvotes (NBR) : Reply Trees Shown & -0.4827727 & 0.0119230 & -40.49 & $< 2\times 10 ^ {-16}$*** \\
Pred. Upvotes (XGB) : Reply Trees Shown & -0.3651081 & 0.0117732 & -31.01 & $< 2\times 10 ^ {-16}$*** \\
---\textit{Editors' Picks Interactions} & & & & \\
Downvotes : Editors' Picks Pinned & -0.0883225 & 0.0095573 & -9.24 & $< 2\times 10 ^ {-16}$*** \\
Rev. Downvotes : Editors' Picks Pinned & -0.0248533 & 0.0095431 & -2.60 & $0.00921$** \\
Upvotes : Editors' Picks Pinned & -0.1553608 & 0.0095800 & -16.22 & $< 2\times 10 ^ {-16}$*** \\
Relative Votes : Editors' Picks Pinned & -0.1500063 & 0.0095768 & -15.66 & $< 2\times 10 ^ {-16}$*** \\
Rev. Chronological : Editors' Picks Pinned & -0.0469577 & 0.0095389 & -4.92 & $8.53\times 10 ^ {-07}$*** \\
Chronological : Editors' Picks Pinned & 0.0749070 & 0.0095314 & 7.86 & $3.87\times 10 ^ {-15}$*** \\
Pred. Editors' Picks (LR) : Editors' Picks Pinned & -0.1999702 & 0.0096582 & -20.70 & $< 2\times 10 ^ {-16}$*** \\
Pred. Editors' Picks (XGB) : Editors' Picks Pinned & -0.2019005 & 0.0096723 & -20.87 & $< 2\times 10 ^ {-16}$*** \\
Pred. Upvotes (NBR) : Editors' Picks Pinned & -0.2314848 & 0.0097891 & -23.65 & $< 2\times 10 ^ {-16}$*** \\
Pred. Upvotes (XGB) : Editors' Picks Pinned & -0.1753108 & 0.0096466 & -18.17 & $< 2\times 10 ^ {-16}$*** \\
\end{tabular}}
\label{tab:results-lexdiv_10}
\end{table}

\begin{table}[]
\centering
\tbl{Regression output for modelling FORUM score for Lexical Diversity over the full comment discussion (***: $p<0.001$, **: $p<0.01$, *: $p<0.05$, .: $p \ge 0.05$).}{
\begin{tabular}{c|c|c|c|c|c}
Outcome & Sample Size & AIC & BIC & Log-Likelihood & Dispersion ($\phi$) \\ \hline 
$\text{FORUM}^{\text{Lexical Diversity}}_{\text{N}}$ & 172,590 & -538722.1 & -538269.4 & 269406.0 & 93.9 \\
\end{tabular}}

\ 

\ 

\resizebox{\textwidth}{!}{
\begin{tabular}{l|r|r|r|c}
Variable & Estimate & Std. Error & $z$-value & $p$-value \\
\hline
---\textit{Main Effects} & & & & \\
(Intercept) & -0.0009038 & 0.0032805 & -0.28 & $0.78293$. \\
Downvotes & 0.2386145 & 0.0046534 & 51.28 & $< 2\times 10 ^ {-16}$*** \\
Rev. Downvotes & -0.0989083 & 0.0046413 & -21.31 & $< 2\times 10 ^ {-16}$*** \\
Upvotes & 0.3243072 & 0.0046650 & 69.52 & $< 2\times 10 ^ {-16}$*** \\
Relative Votes & 0.1953024 & 0.0046489 & 42.01 & $< 2\times 10 ^ {-16}$*** \\
Rev. Chronological & 0.1036262 & 0.0046428 & 22.32 & $< 2\times 10 ^ {-16}$*** \\
Chronological & -0.0976785 & 0.0046414 & -21.05 & $< 2\times 10 ^ {-16}$*** \\
Pred. Editors' Picks (LR) & 0.4903887 & 0.0046988 & 104.36 & $< 2\times 10 ^ {-16}$*** \\
Pred. Editors' Picks (XGB) & 0.7727854 & 0.0047869 & 161.44 & $< 2\times 10 ^ {-16}$*** \\
Pred. Upvotes (NBR) & 0.5000595 & 0.0047021 & 106.35 & $< 2\times 10 ^ {-16}$*** \\
Pred. Upvotes (XGB) & 0.4893645 & 0.0046981 & 104.16 & $< 2\times 10 ^ {-16}$*** \\
Replies Hidden & 0.2228952 & 0.0040289 & 55.32 & $< 2\times 10 ^ {-16}$*** \\
Reply Trees Shown & 0.0180970 & 0.0040157 & 4.51 & $6.59\times 10 ^ {-06}$*** \\
Editors' Picks Pinned & 0.0207471 & 0.0032860 & 6.31 & $2.72\times 10 ^ {-10}$*** \\
---\textit{Reply Structure Interactions} & & & & \\
Downvotes : Replies Hidden & -0.2028833 & 0.0057104 & -35.53 & $< 2\times 10 ^ {-16}$*** \\
Rev. Downvotes : Replies Hidden & 0.0845464 & 0.0056980 & 14.84 & $< 2\times 10 ^ {-16}$*** \\
Upvotes : Replies Hidden & -0.2724705 & 0.0057199 & -47.64 & $< 2\times 10 ^ {-16}$*** \\
Relative Votes : Replies Hidden & -0.1542726 & 0.0057073 & -27.03 & $< 2\times 10 ^ {-16}$*** \\
Rev. Chronological : Replies Hidden & -0.0906131 & 0.0057010 & -15.89 & $< 2\times 10 ^ {-16}$*** \\
Chronological : Replies Hidden & 0.0756088 & 0.0056973 & 13.27 & $< 2\times 10 ^ {-16}$*** \\
Pred. Editors' Picks (LR) : Replies Hidden & -0.3944038 & 0.0057479 & -68.62 & $< 2\times 10 ^ {-16}$*** \\
Pred. Editors' Picks (XGB) : Replies Hidden & -0.6559823 & 0.0058151 & -112.81 & $< 2\times 10 ^ {-16}$*** \\
Pred. Upvotes (NBR) : Replies Hidden & -0.3637816 & 0.0057542 & -63.22 & $< 2\times 10 ^ {-16}$*** \\
Pred. Upvotes (XGB) : Replies Hidden & -0.3886546 & 0.0057486 & -67.61 & $< 2\times 10 ^ {-16}$*** \\
Downvotes : Reply Trees Shown & -0.1651329 & 0.0056907 & -29.02 & $< 2\times 10 ^ {-16}$*** \\
Rev. Downvotes : Reply Trees Shown & 0.0321817 & 0.0056806 & 5.67 & $1.47\times 10 ^ {-08}$*** \\
Upvotes : Reply Trees Shown & -0.2305664 & 0.0056996 & -40.45 & $< 2\times 10 ^ {-16}$*** \\
Relative Votes : Reply Trees Shown & -0.1309978 & 0.0056869 & -23.03 & $< 2\times 10 ^ {-16}$*** \\
Rev. Chronological : Reply Trees Shown & 0.0049165 & 0.0056844 & 0.86 & $0.38709$. \\
Chronological : Reply Trees Shown & -0.0131834 & 0.0056815 & -2.32 & $0.020320$* \\
Pred. Editors' Picks (LR) : Reply Trees Shown & -0.2827482 & 0.0057299 & -49.35 & $< 2\times 10 ^ {-16}$*** \\
Pred. Editors' Picks (XGB) : Reply Trees Shown & -0.4998060 & 0.0058007 & -86.16 & $< 2\times 10 ^ {-16}$*** \\
Pred. Upvotes (NBR) : Reply Trees Shown & -0.2594782 & 0.0057342 & -45.25 & $< 2\times 10 ^ {-16}$*** \\
Pred. Upvotes (XGB) : Reply Trees Shown & -0.3445583 & 0.0057258 & -60.18 & $< 2\times 10 ^ {-16}$*** \\
---\textit{Editors' Picks Interactions} & & & & \\
Downvotes : Editors' Picks Pinned & -0.0120358 & 0.0046548 & -2.59 & $0.009718$** \\
Rev. Downvotes : Editors' Picks Pinned & 0.0030625 & 0.0046473 & 0.66 & $0.50991$. \\
Upvotes : Editors' Picks Pinned & -0.0175022 & 0.0046602 & -3.76 & $0.000173$*** \\
Relative Votes : Editors' Picks Pinned & -0.0154114 & 0.0046529 & -3.31 & $0.000926$*** \\
Rev. Chronological : Editors' Picks Pinned & 0.0009447 & 0.0046506 & 0.20 & $0.83903$. \\
Chronological : Editors' Picks Pinned & -0.0004095 & 0.0046475 & -0.09 & $0.92978$. \\
Pred. Editors' Picks (LR) : Editors' Picks Pinned & -0.0208674 & 0.0046788 & -4.46 & $8.19\times 10 ^ {-06}$*** \\
Pred. Editors' Picks (XGB) : Editors' Picks Pinned & -0.0209862 & 0.0047153 & -4.45 & $8.56\times 10 ^ {-06}$*** \\
Pred. Upvotes (NBR) : Editors' Picks Pinned & -0.0219527 & 0.0046837 & -4.69 & $2.77\times 10 ^ {-06}$*** \\
Pred. Upvotes (XGB) : Editors' Picks Pinned & -0.0182579 & 0.0046768 & -3.90 & $9.46\times 10 ^ {-05}$*** \\
\end{tabular}}
\label{tab:results-lexdiv_N}
\end{table}

\begin{table}[]
\centering
\tbl{Regression output for modelling FORUM score for Sentiment over the first 10 comments (***: $p<0.001$, **: $p<0.01$, *: $p<0.05$, .: $p \ge 0.05$).}{
\begin{tabular}{c|c|c|c|c|c}
Outcome & Sample Size & AIC & BIC & Log-Likelihood & Dispersion ($\phi$) \\ \hline 
$\text{FORUM}^{\text{Sentiment}}_{\text{10}}$ & 172,590 & -170811.3 & -170358.6 & 85450.6 & 10.1 \\
\end{tabular}}

\ 

\ 

\resizebox{\textwidth}{!}{
\begin{tabular}{l|r|r|r|c}
Variable & Estimate & Std. Error & $z$-value & $p$-value \\
\hline
---\textit{Main Effects} & & & & \\
(Intercept) & -0.160354 & 0.009570 & -16.757 & $< 2\times 10 ^ {-16}$*** \\
Downvotes & -0.182733 & 0.013586 & -13.450 & $< 2\times 10 ^ {-16}$*** \\
Rev. Downvotes & 0.039870 & 0.013527 & 2.947 & $0.003204$** \\
Upvotes & -0.081354 & 0.013551 & -6.004 & $1.93\times 10 ^ {-09}$*** \\
Relative Votes & -0.069416 & 0.013547 & -5.124 & $2.99\times 10 ^ {-07}$*** \\
Rev. Chronological & -0.039191 & 0.013540 & -2.894 & $0.003799$** \\
Chronological & 0.044312 & 0.013525 & 3.276 & $0.001052$** \\
Pred. Editors' Picks (LR) & -0.044441 & 0.013542 & -3.282 & $0.001032$** \\
Pred. Editors' Picks (XGB) & -0.049664 & 0.013544 & -3.667 & $0.000246$*** \\
Pred. Upvotes (NBR) & -0.250109 & 0.013620 & -18.363 & $< 2\times 10 ^ {-16}$*** \\
Pred. Upvotes (XGB) & -0.190256 & 0.013593 & -13.997 & $< 2\times 10 ^ {-16}$*** \\
Replies Hidden & -0.010125 & 0.011723 & -0.864 & $0.38771$. \\
Reply Trees Shown & 0.001539 & 0.011720 & 0.131 & $0.89551$. \\
Editors' Picks Pinned & -0.013399 & 0.009571 & -1.400 & $0.16152$. \\
---\textit{Reply Structure Interactions} & & & & \\
Downvotes : Replies Hidden & 0.005794 & 0.016637 & 0.348 & $0.72763$. \\
Rev. Downvotes : Replies Hidden & 0.019928 & 0.016578 & 1.202 & $0.22934$. \\
Upvotes : Replies Hidden & 0.015538 & 0.016599 & 0.936 & $0.34922$. \\
Relative Votes : Replies Hidden & 0.018534 & 0.016594 & 1.117 & $0.26403$. \\
Rev. Chronological : Replies Hidden & 0.029207 & 0.016584 & 1.761 & $0.078219$. \\
Chronological : Replies Hidden & 0.006964 & 0.016568 & 0.420 & $0.67425$. \\
Pred. Editors' Picks (LR) : Replies Hidden & 0.010018 & 0.016589 & 0.604 & $0.54592$. \\
Pred. Editors' Picks (XGB) : Replies Hidden & 0.007226 & 0.016593 & 0.435 & $0.66322$. \\
Pred. Upvotes (NBR) : Replies Hidden & 0.010227 & 0.016682 & 0.613 & $0.53984$. \\
Pred. Upvotes (XGB) : Replies Hidden & 0.012163 & 0.016652 & 0.730 & $0.46512$. \\
Downvotes : Reply Trees Shown & 0.067823 & 0.016619 & 4.081 & $4.48\times 10 ^ {-05}$*** \\
Rev. Downvotes : Reply Trees Shown & -0.002821 & 0.016575 & -0.170 & $0.86483$. \\
Upvotes : Reply Trees Shown & 0.067270 & 0.016587 & 4.056 & $5.00\times 10 ^ {-05}$*** \\
Relative Votes : Reply Trees Shown & 0.062651 & 0.016584 & 3.778 & $0.000158$*** \\
Rev. Chronological : Reply Trees Shown & 0.031765 & 0.016581 & 1.916 & $0.055391$. \\
Chronological : Reply Trees Shown & -0.006556 & 0.016567 & -0.396 & $0.69232$. \\
Pred. Editors' Picks (LR) : Reply Trees Shown & 0.041513 & 0.016580 & 2.504 & $0.012289$* \\
Pred. Editors' Picks (XGB) : Reply Trees Shown & 0.044257 & 0.016583 & 2.669 & $0.007612$** \\
Pred. Upvotes (NBR) : Reply Trees Shown & 0.134397 & 0.016644 & 8.075 & $6.76\times 10 ^ {-16}$*** \\
Pred. Upvotes (XGB) : Reply Trees Shown & 0.115895 & 0.016623 & 6.972 & $3.13\times 10 ^ {-12}$*** \\
---\textit{Editors' Picks Interactions} & & & & \\
Downvotes : Editors' Picks Pinned & 0.066663 & 0.013575 & 4.911 & $9.07\times 10 ^ {-07}$*** \\
Rev. Downvotes : Editors' Picks Pinned & -0.072721 & 0.013534 & -5.373 & $7.74\times 10 ^ {-08}$*** \\
Upvotes : Editors' Picks Pinned & 0.022201 & 0.013546 & 1.639 & $0.10124$. \\
Relative Votes : Editors' Picks Pinned & 0.019274 & 0.013544 & 1.423 & $0.15471$. \\
Rev. Chronological : Editors' Picks Pinned & 0.007739 & 0.013539 & 0.572 & $0.56759$. \\
Chronological : Editors' Picks Pinned & -0.015088 & 0.013528 & -1.115 & $0.26470$. \\
Pred. Editors' Picks (LR) : Editors' Picks Pinned & 0.010988 & 0.013540 & 0.811 & $0.41709$. \\
Pred. Editors' Picks (XGB) : Editors' Picks Pinned & 0.001485 & 0.013543 & 0.110 & $0.91266$. \\
Pred. Upvotes (NBR) : Editors' Picks Pinned & 0.042843 & 0.013600 & 3.150 & $0.001632$** \\
Pred. Upvotes (XGB) : Editors' Picks Pinned & 0.020551 & 0.013581 & 1.513 & $0.13021$. \\
\end{tabular}}
\label{tab:results-sentcomp_10}
\end{table}

\begin{table}[]
\centering
\tbl{Regression output for modelling FORUM score for Sentiment over the full comment discussion (***: $p<0.001$, **: $p<0.01$, *: $p<0.05$, .: $p \ge 0.05$).}{
\begin{tabular}{c|c|c|c|c|c}
Outcome & Sample Size & AIC & BIC & Log-Likelihood & Dispersion ($\phi$) \\ \hline 
$\text{FORUM}^{\text{Sentiment}}_{\text{N}}$ & 172,590 & -577640.2 & -577187.5 & 288865.1 & 120 \\
\end{tabular}}

\ 

\ 

\resizebox{\textwidth}{!}{
\begin{tabular}{l|r|r|r|c}
Variable & Estimate & Std. Error & $z$-value & $p$-value \\
\hline
---\textit{Main Effects} & & & & \\
(Intercept) & 2.823e-04 & 2.899e-03 & 0.097 & $0.92243$. \\
Downvotes & -8.195e-02 & 4.102e-03 & -19.980 & $< 2\times 10 ^ {-16}$*** \\
Rev. Downvotes & 8.651e-02 & 4.102e-03 & 21.089 & $< 2\times 10 ^ {-16}$*** \\
Upvotes & -3.010e-02 & 4.100e-03 & -7.342 & $2.10\times 10 ^ {-13}$*** \\
Relative Votes & 1.583e-02 & 4.100e-03 & 3.861 & $0.000113$*** \\
Rev. Chronological & -3.602e-02 & 4.100e-03 & -8.784 & $< 2\times 10 ^ {-16}$*** \\
Chronological & 3.584e-02 & 4.100e-03 & 8.740 & $< 2\times 10 ^ {-16}$*** \\
Pred. Editors' Picks (LR) & -3.446e-02 & 4.100e-03 & -8.405 & $< 2\times 10 ^ {-16}$*** \\
Pred. Editors' Picks (XGB) & -1.070e-01 & 4.103e-03 & -26.081 & $< 2\times 10 ^ {-16}$*** \\
Pred. Upvotes (NBR) & -5.817e-02 & 4.101e-03 & -14.183 & $< 2\times 10 ^ {-16}$*** \\
Pred. Upvotes (XGB) & -4.549e-02 & 4.101e-03 & -11.094 & $< 2\times 10 ^ {-16}$*** \\
Replies Hidden & -1.698e-02 & 3.551e-03 & -4.783 & $1.73\times 10 ^ {-06}$*** \\
Reply Trees Shown & -5.607e-03 & 3.551e-03 & -1.579 & $0.11429$. \\
Editors' Picks Pinned & 9.912e-04 & 2.899e-03 & 0.342 & $0.73244$. \\
---\textit{Reply Structure Interactions} & & & & \\
Downvotes : Replies Hidden & 5.783e-02 & 5.023e-03 & 11.514 & $< 2\times 10 ^ {-16}$*** \\
Rev. Downvotes : Replies Hidden & -5.232e-02 & 5.023e-03 & -10.417 & $< 2\times 10 ^ {-16}$*** \\
Upvotes : Replies Hidden & 1.837e-02 & 5.022e-03 & 3.657 & $0.000255$*** \\
Relative Votes : Replies Hidden & -2.001e-02 & 5.022e-03 & -3.985 & $6.74\times 10 ^ {-05}$*** \\
Rev. Chronological : Replies Hidden & 3.289e-02 & 5.022e-03 & 6.549 & $5.81\times 10 ^ {-11}$*** \\
Chronological : Replies Hidden & -2.807e-02 & 5.022e-03 & -5.590 & $2.27\times 10 ^ {-08}$*** \\
Pred. Editors' Picks (LR) : Replies Hidden & 3.338e-02 & 5.022e-03 & 6.648 & $2.98\times 10 ^ {-11}$*** \\
Pred. Editors' Picks (XGB) : Replies Hidden & 8.649e-02 & 5.024e-03 & 17.217 & $< 2\times 10 ^ {-16}$*** \\
Pred. Upvotes (NBR) : Replies Hidden & 1.882e-02 & 5.023e-03 & 3.747 & $0.000179$*** \\
Pred. Upvotes (XGB) : Replies Hidden & 1.337e-02 & 5.022e-03 & 2.663 & $0.007749$** \\
Downvotes : Reply Trees Shown & 1.688e-02 & 5.023e-03 & 3.360 & $0.000778$*** \\
Rev. Downvotes : Reply Trees Shown & -5.395e-02 & 5.022e-03 & -10.743 & $< 2\times 10 ^ {-16}$*** \\
Upvotes : Reply Trees Shown & 1.759e-02 & 5.022e-03 & 3.503 & $0.000460$*** \\
Relative Votes : Reply Trees Shown & 8.292e-03 & 5.022e-03 & 1.651 & $0.098684$. \\
Rev. Chronological : Reply Trees Shown & 2.201e-02 & 5.022e-03 & 4.384 & $1.17\times 10 ^ {-05}$*** \\
Chronological : Reply Trees Shown & -1.510e-02 & 5.022e-03 & -3.008 & $0.002633$** \\
Pred. Editors' Picks (LR) : Reply Trees Shown & 3.337e-02 & 5.022e-03 & 6.645 & $3.03\times 10 ^ {-11}$*** \\
Pred. Editors' Picks (XGB) : Reply Trees Shown & 4.774e-02 & 5.024e-03 & 9.503 & $< 2\times 10 ^ {-16}$*** \\
Pred. Upvotes (NBR) : Reply Trees Shown & -2.333e-02 & 5.023e-03 & -4.645 & $3.41\times 10 ^ {-06}$*** \\
Pred. Upvotes (XGB) : Reply Trees Shown & 5.834e-03 & 5.022e-03 & 1.162 & $0.24541$. \\
---\textit{Editors' Picks Interactions} & & & & \\
Downvotes : Editors' Picks Pinned & 3.532e-03 & 4.101e-03 & 0.861 & $0.38915$. \\
Rev. Downvotes : Editors' Picks Pinned & -4.231e-02 & 4.101e-03 & -10.317 & $< 2\times 10 ^ {-16}$*** \\
Upvotes : Editors' Picks Pinned & -6.696e-04 & 4.100e-03 & -0.163 & $0.87028$. \\
Relative Votes : Editors' Picks Pinned & -1.393e-03 & 4.100e-03 & -0.340 & $0.73413$. \\
Rev. Chronological : Editors' Picks Pinned & -5.841e-04 & 4.100e-03 & -0.142 & $0.88672$. \\
Chronological : Editors' Picks Pinned & -1.668e-03 & 4.100e-03 & -0.407 & $0.68423$. \\
Pred. Editors' Picks (LR) : Editors' Picks Pinned & -1.319e-03 & 4.100e-03 & -0.322 & $0.74768$. \\
Pred. Editors' Picks (XGB) : Editors' Picks Pinned & -1.516e-03 & 4.102e-03 & -0.370 & $0.71171$. \\
Pred. Upvotes (NBR) : Editors' Picks Pinned & 4.333e-05 & 4.101e-03 & 0.011 & $0.99157$. \\
Pred. Upvotes (XGB) : Editors' Picks Pinned & -1.010e-03 & 4.101e-03 & -0.246 & $0.80542$. \\
\end{tabular}}
\label{tab:results-sentcomp_N}
\end{table}

\begin{table}[]
\centering
\tbl{Regression output for modelling FORUM score for Topical Similarity over the first 10 comments (***: $p<0.001$, **: $p<0.01$, *: $p<0.05$, .: $p \ge 0.05$).}{
\begin{tabular}{c|c|c|c|c|c}
Outcome & Sample Size & AIC & BIC & Log-Likelihood & Dispersion ($\phi$) \\ \hline 
$\text{FORUM}^{\text{Topical Similarity}}_{\text{10}}$ & 172,590 & -153415.8 & -152963.1 & 76752.9 & 9.02 \\
\end{tabular}}

\ 

\ 

\resizebox{\textwidth}{!}{
\begin{tabular}{l|r|r|r|c}
Variable & Estimate & Std. Error & $z$-value & $p$-value \\
\hline
---\textit{Main Effects} & & & & \\
(Intercept) & -0.3662245 & 0.0101289 & -36.16 & $< 2\times 10 ^ {-16}$*** \\
Downvotes & 0.2319087 & 0.0142901 & 16.23 & $< 2\times 10 ^ {-16}$*** \\
Rev. Downvotes & 0.2926566 & 0.0142822 & 20.49 & $< 2\times 10 ^ {-16}$*** \\
Upvotes & 0.4461850 & 0.0142807 & 31.24 & $< 2\times 10 ^ {-16}$*** \\
Relative Votes & 0.4459127 & 0.0142804 & 31.23 & $< 2\times 10 ^ {-16}$*** \\
Rev. Chronological & -0.1362846 & 0.0143581 & -9.49 & $< 2\times 10 ^ {-16}$*** \\
Chronological & 0.1755314 & 0.0142895 & 12.28 & $< 2\times 10 ^ {-16}$*** \\
Pred. Editors' Picks (LR) & 0.5363823 & 0.0142941 & 37.52 & $< 2\times 10 ^ {-16}$*** \\
Pred. Editors' Picks (XGB) & 0.5248869 & 0.0142913 & 36.73 & $< 2\times 10 ^ {-16}$*** \\
Pred. Upvotes (NBR) & 0.6166586 & 0.0143147 & 43.08 & $< 2\times 10 ^ {-16}$*** \\
Pred. Upvotes (XGB) & 0.5424462 & 0.0142967 & 37.94 & $< 2\times 10 ^ {-16}$*** \\
Replies Hidden & 0.3594616 & 0.0123748 & 29.05 & $< 2\times 10 ^ {-16}$*** \\
Reply Trees Shown & 0.0377722 & 0.0124200 & 3.04 & $0.00236$** \\
Editors' Picks Pinned & 0.0924426 & 0.0101139 & 9.14 & $< 2\times 10 ^ {-16}$*** \\
---\textit{Reply Structure Interactions} & & & & \\
Downvotes : Replies Hidden & -0.2361534 & 0.0174624 & -13.52 & $< 2\times 10 ^ {-16}$*** \\
Rev. Downvotes : Replies Hidden & -0.2218157 & 0.0174611 & -12.70 & $< 2\times 10 ^ {-16}$*** \\
Upvotes : Replies Hidden & -0.3086634 & 0.0174675 & -17.67 & $< 2\times 10 ^ {-16}$*** \\
Relative Votes : Replies Hidden & -0.3029843 & 0.0174680 & -17.35 & $< 2\times 10 ^ {-16}$*** \\
Rev. Chronological : Replies Hidden & 0.1550237 & 0.0175419 & 8.84 & $< 2\times 10 ^ {-16}$*** \\
Chronological : Replies Hidden & -0.2332848 & 0.0174656 & -13.36 & $< 2\times 10 ^ {-16}$*** \\
Pred. Editors' Picks (LR) : Replies Hidden & -0.3594395 & 0.0174794 & -20.56 & $< 2\times 10 ^ {-16}$*** \\
Pred. Editors' Picks (XGB) : Replies Hidden & -0.3554489 & 0.0174765 & -20.34 & $< 2\times 10 ^ {-16}$*** \\
Pred. Upvotes (NBR) : Replies Hidden & -0.3593726 & 0.0175062 & -20.53 & $< 2\times 10 ^ {-16}$*** \\
Pred. Upvotes (XGB) : Replies Hidden & -0.3564016 & 0.0174836 & -20.38 & $< 2\times 10 ^ {-16}$*** \\
Downvotes : Reply Trees Shown & -0.3920579 & 0.0175770 & -22.31 & $< 2\times 10 ^ {-16}$*** \\
Rev. Downvotes : Reply Trees Shown & -0.3989832 & 0.0175733 & -22.70 & $< 2\times 10 ^ {-16}$*** \\
Upvotes : Reply Trees Shown & -0.4081708 & 0.0175261 & -23.29 & $< 2\times 10 ^ {-16}$*** \\
Relative Votes : Reply Trees Shown & -0.3914340 & 0.0175223 & -22.34 & $< 2\times 10 ^ {-16}$*** \\
Rev. Chronological : Reply Trees Shown & 0.2917259 & 0.0175833 & 16.59 & $< 2\times 10 ^ {-16}$*** \\
Chronological : Reply Trees Shown & -0.2470110 & 0.0175464 & -14.08 & $< 2\times 10 ^ {-16}$*** \\
Pred. Editors' Picks (LR) : Reply Trees Shown & -0.4707094 & 0.0175307 & -26.85 & $< 2\times 10 ^ {-16}$*** \\
Pred. Editors' Picks (XGB) : Reply Trees Shown & -0.4487201 & 0.0175270 & -25.60 & $< 2\times 10 ^ {-16}$*** \\
Pred. Upvotes (NBR) : Reply Trees Shown & -0.5583287 & 0.0175465 & -31.82 & $< 2\times 10 ^ {-16}$*** \\
Pred. Upvotes (XGB) : Reply Trees Shown & -0.5333004 & 0.0175448 & -30.40 & $< 2\times 10 ^ {-16}$*** \\
---\textit{Editors' Picks Interactions} & & & & \\
Downvotes : Editors' Picks Pinned & -0.0114123 & 0.0143080 & -0.80 & $0.4250$. \\
Rev. Downvotes : Editors' Picks Pinned & -0.1068475 & 0.0143062 & -7.47 & $8.10\times 10 ^ {-14}$*** \\
Upvotes : Editors' Picks Pinned & -0.0799877 & 0.0142845 & -5.60 & $2.15\times 10 ^ {-08}$*** \\
Relative Votes : Editors' Picks Pinned & -0.0806808 & 0.0142828 & -5.65 & $1.62\times 10 ^ {-08}$*** \\
Rev. Chronological : Editors' Picks Pinned & 0.0002806 & 0.0143126 & 0.02 & $0.9843$. \\
Chronological : Editors' Picks Pinned & 0.0263122 & 0.0142926 & 1.84 & $0.06563$. \\
Pred. Editors' Picks (LR) : Editors' Picks Pinned & -0.1025777 & 0.0142889 & -7.18 & $7.03\times 10 ^ {-13}$*** \\
Pred. Editors' Picks (XGB) : Editors' Picks Pinned & -0.1056971 & 0.0142863 & -7.40 & $1.38\times 10 ^ {-13}$*** \\
Pred. Upvotes (NBR) : Editors' Picks Pinned & -0.1057021 & 0.0143046 & -7.39 & $1.48\times 10 ^ {-13}$*** \\
Pred. Upvotes (XGB) : Editors' Picks Pinned & -0.0880212 & 0.0142976 & -6.16 & $7.44\times 10 ^ {-10}$*** \\
\end{tabular}}
\label{tab:results-sim_10}
\end{table}

\begin{table}[]
\centering
\tbl{Regression output for modelling FORUM score for Topical Similarity over the full comment discussion (***: $p<0.001$, **: $p<0.01$, *: $p<0.05$, .: $p \ge 0.05$).}{
\begin{tabular}{c|c|c|c|c|c}
Outcome & Sample Size & AIC & BIC & Log-Likelihood & Dispersion ($\phi$) \\ \hline 
$\text{FORUM}^{\text{Topical Similarity}}_{\text{N}}$ & 172,590 & -509572.7 & -509120.0 & 254831.3 & 77.6 \\
\end{tabular}}

\ 

\ 

\resizebox{\textwidth}{!}{
\begin{tabular}{l|r|r|r|c}
Variable & Estimate & Std. Error & $z$-value & $p$-value \\
\hline
---\textit{Main Effects} & & & & \\
(Intercept) & 0.0041385 & 0.0036109 & 1.15 & $0.2517$. \\
Downvotes & 0.2118989 & 0.0051192 & 41.39 & $< 2\times 10 ^ {-16}$*** \\
Rev. Downvotes & -0.0148000 & 0.0051078 & -2.90 & $0.00376$** \\
Upvotes & 0.4313026 & 0.0051566 & 83.64 & $< 2\times 10 ^ {-16}$*** \\
Relative Votes & 0.2809940 & 0.0051285 & 54.79 & $< 2\times 10 ^ {-16}$*** \\
Rev. Chronological & -0.1253356 & 0.0051100 & -24.53 & $< 2\times 10 ^ {-16}$*** \\
Chronological & 0.1356236 & 0.0051119 & 26.53 & $< 2\times 10 ^ {-16}$*** \\
Pred. Editors' Picks (LR) & 0.5543683 & 0.0051892 & 106.83 & $< 2\times 10 ^ {-16}$*** \\
Pred. Editors' Picks (XGB) & 0.7138516 & 0.0052435 & 136.14 & $< 2\times 10 ^ {-16}$*** \\
Pred. Upvotes (NBR) & 0.7190192 & 0.0052468 & 137.04 & $< 2\times 10 ^ {-16}$*** \\
Pred. Upvotes (XGB) & 0.6722930 & 0.0052286 & 128.58 & $< 2\times 10 ^ {-16}$*** \\
Replies Hidden & 0.4896618 & 0.0044784 & 109.34 & $< 2\times 10 ^ {-16}$*** \\
Reply Trees Shown & 0.0081949 & 0.0044117 & 1.86 & $0.06323$. \\
Editors' Picks Pinned & 0.0039217 & 0.0036371 & 1.08 & $0.2809$. \\
---\textit{Reply Structure Interactions} & & & & \\
Downvotes : Replies Hidden & -0.2142980 & 0.0063424 & -33.79 & $< 2\times 10 ^ {-16}$*** \\
Rev. Downvotes : Replies Hidden & 0.0723230 & 0.0063382 & 11.41 & $< 2\times 10 ^ {-16}$*** \\
Upvotes : Replies Hidden & -0.4082599 & 0.0063737 & -64.05 & $< 2\times 10 ^ {-16}$*** \\
Relative Votes : Replies Hidden & -0.2573971 & 0.0063529 & -40.52 & $< 2\times 10 ^ {-16}$*** \\
Rev. Chronological : Replies Hidden & 0.1271659 & 0.0063363 & 20.07 & $< 2\times 10 ^ {-16}$*** \\
Chronological : Replies Hidden & -0.1398733 & 0.0063365 & -22.07 & $< 2\times 10 ^ {-16}$*** \\
Pred. Editors' Picks (LR) : Replies Hidden & -0.5214065 & 0.0063990 & -81.48 & $< 2\times 10 ^ {-16}$*** \\
Pred. Editors' Picks (XGB) : Replies Hidden & -0.6854887 & 0.0064379 & -106.48 & $< 2\times 10 ^ {-16}$*** \\
Pred. Upvotes (NBR) : Replies Hidden & -0.6608280 & 0.0064458 & -102.52 & $< 2\times 10 ^ {-16}$*** \\
Pred. Upvotes (XGB) : Replies Hidden & -0.6251012 & 0.0064307 & -97.21 & $< 2\times 10 ^ {-16}$*** \\
Downvotes : Reply Trees Shown & -0.3570286 & 0.0062514 & -57.11 & $< 2\times 10 ^ {-16}$*** \\
Rev. Downvotes : Reply Trees Shown & 0.0684302 & 0.0062398 & 10.97 & $< 2\times 10 ^ {-16}$*** \\
Upvotes : Reply Trees Shown & -0.4885424 & 0.0062760 & -77.84 & $< 2\times 10 ^ {-16}$*** \\
Relative Votes : Reply Trees Shown & -0.2213501 & 0.0062555 & -35.38 & $< 2\times 10 ^ {-16}$*** \\
Rev. Chronological : Reply Trees Shown & 0.2088894 & 0.0062435 & 33.46 & $< 2\times 10 ^ {-16}$*** \\
Chronological : Reply Trees Shown & -0.2262408 & 0.0062439 & -36.23 & $< 2\times 10 ^ {-16}$*** \\
Pred. Editors' Picks (LR) : Reply Trees Shown & -0.6414037 & 0.0063005 & -101.80 & $< 2\times 10 ^ {-16}$*** \\
Pred. Editors' Picks (XGB) : Reply Trees Shown & -0.8298095 & 0.0063420 & -130.84 & $< 2\times 10 ^ {-16}$*** \\
Pred. Upvotes (NBR) : Reply Trees Shown & -0.9054729 & 0.0063473 & -142.65 & $< 2\times 10 ^ {-16}$*** \\
Pred. Upvotes (XGB) : Reply Trees Shown & -0.7849854 & 0.0063304 & -124.00 & $< 2\times 10 ^ {-16}$*** \\
---\textit{Editors' Picks Interactions} & & & & \\
Downvotes : Editors' Picks Pinned & 0.0039707 & 0.0051505 & 0.77 & $0.4407$. \\
Rev. Downvotes : Editors' Picks Pinned & -0.1108737 & 0.0051454 & -21.55 & $< 2\times 10 ^ {-16}$*** \\
Upvotes : Editors' Picks Pinned & -0.0022882 & 0.0051660 & -0.44 & $0.6578$. \\
Relative Votes : Editors' Picks Pinned & -0.0036942 & 0.0051550 & -0.72 & $0.4736$. \\
Rev. Chronological : Editors' Picks Pinned & -0.0006544 & 0.0051463 & -0.13 & $0.8988$. \\
Chronological : Editors' Picks Pinned & 0.0012145 & 0.0051460 & 0.24 & $0.8134$. \\
Pred. Editors' Picks (LR) : Editors' Picks Pinned & -0.0039219 & 0.0051797 & -0.76 & $0.4489$. \\
Pred. Editors' Picks (XGB) : Editors' Picks Pinned & -0.0046698 & 0.0051998 & -0.90 & $0.3691$. \\
Pred. Upvotes (NBR) : Editors' Picks Pinned & -0.0038031 & 0.0052063 & -0.73 & $0.4651$. \\
Pred. Upvotes (XGB) : Editors' Picks Pinned & -0.0029638 & 0.0051964 & -0.57 & $0.5684$. \\
\end{tabular}}
\label{tab:results-sim_N}
\end{table}

\begin{table}[]
\centering
\tbl{Regression output for modelling FORUM score for Readability over the first 10 comments (***: $p<0.001$, **: $p<0.01$, *: $p<0.05$, .: $p \ge 0.05$).}{
\begin{tabular}{c|c|c|c|c|c}
Outcome & Sample Size & AIC & BIC & Log-Likelihood & Dispersion ($\phi$) \\ \hline 
$\text{FORUM}^{\text{Readability}}_{\text{10}}$ & 172,590 & -325216.6 & -324764.0 & 162653.3 & 26.7 \\
\end{tabular}}

\ 

\ 

\resizebox{\textwidth}{!}{
\begin{tabular}{l|r|r|r|c}
Variable & Estimate & Std. Error & $z$-value & $p$-value \\
\hline
---\textit{Main Effects} & & & & \\
(Intercept) & -0.076128 & 0.006064 & -12.55 & $< 2\times 10 ^ {-16}$*** \\
Downvotes & 0.159646 & 0.008580 & 18.61 & $< 2\times 10 ^ {-16}$*** \\
Rev. Downvotes & 0.070754 & 0.008576 & 8.25 & $< 2\times 10 ^ {-16}$*** \\
Upvotes & 0.276182 & 0.008593 & 32.14 & $< 2\times 10 ^ {-16}$*** \\
Relative Votes & 0.270111 & 0.008592 & 31.44 & $< 2\times 10 ^ {-16}$*** \\
Rev. Chronological & -0.022245 & 0.008576 & -2.59 & $0.009490$** \\
Chronological & -0.080687 & 0.008582 & -9.40 & $< 2\times 10 ^ {-16}$*** \\
Pred. Editors' Picks (LR) & 0.383883 & 0.008618 & 44.55 & $< 2\times 10 ^ {-16}$*** \\
Pred. Editors' Picks (XGB) & 0.413056 & 0.008626 & 47.88 & $< 2\times 10 ^ {-16}$*** \\
Pred. Upvotes (NBR) & 0.555167 & 0.008678 & 63.97 & $< 2\times 10 ^ {-16}$*** \\
Pred. Upvotes (XGB) & 0.391229 & 0.008620 & 45.39 & $< 2\times 10 ^ {-16}$*** \\
Replies Hidden & 0.114836 & 0.007429 & 15.46 & $< 2\times 10 ^ {-16}$*** \\
Reply Trees Shown & -0.002913 & 0.007427 & -0.39 & $0.69486$. \\
Editors' Picks Pinned & 0.104163 & 0.006066 & 17.17 & $< 2\times 10 ^ {-16}$*** \\
---\textit{Reply Structure Interactions} & & & & \\
Downvotes : Replies Hidden & -0.070789 & 0.010515 & -6.73 & $1.67\times 10 ^ {-11}$*** \\
Rev. Downvotes : Replies Hidden & -0.077147 & 0.010508 & -7.34 & $2.10\times 10 ^ {-13}$*** \\
Upvotes : Replies Hidden & -0.100235 & 0.010531 & -9.52 & $< 2\times 10 ^ {-16}$*** \\
Relative Votes : Replies Hidden & -0.098945 & 0.010530 & -9.40 & $< 2\times 10 ^ {-16}$*** \\
Rev. Chronological : Replies Hidden & 0.035577 & 0.010507 & 3.39 & $0.000709$*** \\
Chronological : Replies Hidden & -0.056593 & 0.010509 & -5.39 & $7.23\times 10 ^ {-08}$*** \\
Pred. Editors' Picks (LR) : Replies Hidden & -0.114807 & 0.010560 & -10.87 & $< 2\times 10 ^ {-16}$*** \\
Pred. Editors' Picks (XGB) : Replies Hidden & -0.117853 & 0.010570 & -11.15 & $< 2\times 10 ^ {-16}$*** \\
Pred. Upvotes (NBR) : Replies Hidden & -0.115037 & 0.010636 & -10.82 & $< 2\times 10 ^ {-16}$*** \\
Pred. Upvotes (XGB) : Replies Hidden & -0.115193 & 0.010564 & -10.90 & $< 2\times 10 ^ {-16}$*** \\
Downvotes : Reply Trees Shown & -0.156245 & 0.010507 & -14.87 & $< 2\times 10 ^ {-16}$*** \\
Rev. Downvotes : Reply Trees Shown & -0.198389 & 0.010511 & -18.87 & $< 2\times 10 ^ {-16}$*** \\
Upvotes : Reply Trees Shown & -0.172068 & 0.010515 & -16.36 & $< 2\times 10 ^ {-16}$*** \\
Relative Votes : Reply Trees Shown & -0.170537 & 0.010514 & -16.22 & $< 2\times 10 ^ {-16}$*** \\
Rev. Chronological : Reply Trees Shown & 0.075017 & 0.010504 & 7.14 & $9.20\times 10 ^ {-13}$*** \\
Chronological : Reply Trees Shown & -0.052031 & 0.010512 & -4.95 & $7.44\times 10 ^ {-07}$*** \\
Pred. Editors' Picks (LR) : Reply Trees Shown & -0.213580 & 0.010532 & -20.28 & $< 2\times 10 ^ {-16}$*** \\
Pred. Editors' Picks (XGB) : Reply Trees Shown & -0.228993 & 0.010539 & -21.73 & $< 2\times 10 ^ {-16}$*** \\
Pred. Upvotes (NBR) : Reply Trees Shown & -0.306665 & 0.010576 & -29.00 & $< 2\times 10 ^ {-16}$*** \\
Pred. Upvotes (XGB) : Reply Trees Shown & -0.245199 & 0.010534 & -23.28 & $< 2\times 10 ^ {-16}$*** \\
---\textit{Editors' Picks Interactions} & & & & \\
Downvotes : Editors' Picks Pinned & -0.042799 & 0.008583 & -4.99 & $6.15\times 10 ^ {-07}$*** \\
Rev. Downvotes : Editors' Picks Pinned & -0.009805 & 0.008582 & -1.14 & $0.25328$. \\
Upvotes : Editors' Picks Pinned & -0.090354 & 0.008591 & -10.52 & $< 2\times 10 ^ {-16}$*** \\
Relative Votes : Editors' Picks Pinned & -0.088781 & 0.008590 & -10.33 & $< 2\times 10 ^ {-16}$*** \\
Rev. Chronological : Editors' Picks Pinned & -0.019662 & 0.008579 & -2.29 & $0.021912$* \\
Chronological : Editors' Picks Pinned & 0.060642 & 0.008583 & 7.07 & $1.60\times 10 ^ {-12}$*** \\
Pred. Editors' Picks (LR) : Editors' Picks Pinned & -0.109832 & 0.008608 & -12.76 & $< 2\times 10 ^ {-16}$*** \\
Pred. Editors' Picks (XGB) : Editors' Picks Pinned & -0.109904 & 0.008614 & -12.76 & $< 2\times 10 ^ {-16}$*** \\
Pred. Upvotes (NBR) : Editors' Picks Pinned & -0.128739 & 0.008652 & -14.88 & $< 2\times 10 ^ {-16}$*** \\
Pred. Upvotes (XGB) : Editors' Picks Pinned & -0.100056 & 0.008610 & -11.62 & $< 2\times 10 ^ {-16}$*** \\
\end{tabular}}
\label{tab:results-smog_10}
\end{table}

\begin{table}[]
\centering
\tbl{Regression output for modelling FORUM score for Readability over the full comment discussion (***: $p<0.001$, **: $p<0.01$, *: $p<0.05$, .: $p \ge 0.05$).}{
\begin{tabular}{c|c|c|c|c|c}
Outcome & Sample Size & AIC & BIC & Log-Likelihood & Dispersion ($\phi$) \\ \hline 
$\text{FORUM}^{\text{Readability}}_{\text{N}}$ & 172,590 & -582018.6 & -581566.0 & 291054.3 & 122 \\
\end{tabular}}

\ 

\ 

\resizebox{\textwidth}{!}{
\begin{tabular}{l|r|r|r|c}
Variable & Estimate & Std. Error & $z$-value & $p$-value \\
\hline
---\textit{Main Effects} & & & & \\
(Intercept) & -0.0006671 & 0.0028791 & -0.23 & $0.81675$. \\
Downvotes & 0.1655857 & 0.0040775 & 40.61 & $< 2\times 10 ^ {-16}$*** \\
Rev. Downvotes & -0.0735209 & 0.0040726 & -18.05 & $< 2\times 10 ^ {-16}$*** \\
Upvotes & 0.2594346 & 0.0040860 & 63.49 & $< 2\times 10 ^ {-16}$*** \\
Relative Votes & 0.1692625 & 0.0040778 & 41.51 & $< 2\times 10 ^ {-16}$*** \\
Rev. Chronological & 0.0432267 & 0.0040723 & 10.61 & $< 2\times 10 ^ {-16}$*** \\
Chronological & -0.0388535 & 0.0040718 & -9.54 & $< 2\times 10 ^ {-16}$*** \\
Pred. Editors' Picks (LR) & 0.3448901 & 0.0040973 & 84.17 & $< 2\times 10 ^ {-16}$*** \\
Pred. Editors' Picks (XGB) & 0.5220685 & 0.0041302 & 126.40 & $< 2\times 10 ^ {-16}$*** \\
Pred. Upvotes (NBR) & 0.4253794 & 0.0041111 & 103.47 & $< 2\times 10 ^ {-16}$*** \\
Pred. Upvotes (XGB) & 0.3977503 & 0.0041056 & 96.88 & $< 2\times 10 ^ {-16}$*** \\
Replies Hidden & 0.1654455 & 0.0035315 & 46.85 & $< 2\times 10 ^ {-16}$*** \\
Reply Trees Shown & 0.0134479 & 0.0035251 & 3.81 & $0.000136$*** \\
Editors' Picks Pinned & 0.0145081 & 0.0028817 & 5.03 & $4.79\times 10 ^ {-07}$*** \\
---\textit{Reply Structure Interactions} & & & & \\
Downvotes : Replies Hidden & -0.1399405 & 0.0049997 & -27.99 & $< 2\times 10 ^ {-16}$*** \\
Rev. Downvotes : Replies Hidden & 0.0679022 & 0.0049946 & 13.60 & $< 2\times 10 ^ {-16}$*** \\
Upvotes : Replies Hidden & -0.2197128 & 0.0050065 & -43.89 & $< 2\times 10 ^ {-16}$*** \\
Relative Votes : Replies Hidden & -0.1363686 & 0.0050002 & -27.27 & $< 2\times 10 ^ {-16}$*** \\
Rev. Chronological : Replies Hidden & -0.0343188 & 0.0049951 & -6.87 & $6.40\times 10 ^ {-12}$*** \\
Chronological : Replies Hidden & 0.0289879 & 0.0049938 & 5.80 & $6.45\times 10 ^ {-09}$*** \\
Pred. Editors' Picks (LR) : Replies Hidden & -0.2794664 & 0.0050161 & -55.71 & $< 2\times 10 ^ {-16}$*** \\
Pred. Editors' Picks (XGB) : Replies Hidden & -0.4440744 & 0.0050410 & -88.09 & $< 2\times 10 ^ {-16}$*** \\
Pred. Upvotes (NBR) : Replies Hidden & -0.3201475 & 0.0050287 & -63.66 & $< 2\times 10 ^ {-16}$*** \\
Pred. Upvotes (XGB) : Replies Hidden & -0.3212436 & 0.0050231 & -63.95 & $< 2\times 10 ^ {-16}$*** \\
Downvotes : Reply Trees Shown & -0.1275741 & 0.0049899 & -25.57 & $< 2\times 10 ^ {-16}$*** \\
Rev. Downvotes : Reply Trees Shown & 0.0340916 & 0.0049860 & 6.84 & $8.06\times 10 ^ {-12}$*** \\
Upvotes : Reply Trees Shown & -0.1970249 & 0.0049964 & -39.43 & $< 2\times 10 ^ {-16}$*** \\
Relative Votes : Reply Trees Shown & -0.1208212 & 0.0049902 & -24.21 & $< 2\times 10 ^ {-16}$*** \\
Rev. Chronological : Reply Trees Shown & 0.0209506 & 0.0049866 & 4.20 & $2.65\times 10 ^ {-05}$*** \\
Chronological : Reply Trees Shown & -0.0285466 & 0.0049857 & -5.73 & $1.03\times 10 ^ {-08}$*** \\
Pred. Editors' Picks (LR) : Reply Trees Shown & -0.2159409 & 0.0050066 & -43.13 & $< 2\times 10 ^ {-16}$*** \\
Pred. Editors' Picks (XGB) : Reply Trees Shown & -0.3498950 & 0.0050328 & -69.52 & $< 2\times 10 ^ {-16}$*** \\
Pred. Upvotes (NBR) : Reply Trees Shown & -0.2486296 & 0.0050186 & -49.54 & $< 2\times 10 ^ {-16}$*** \\
Pred. Upvotes (XGB) : Reply Trees Shown & -0.2953501 & 0.0050117 & -58.93 & $< 2\times 10 ^ {-16}$*** \\
---\textit{Editors' Picks Interactions} & & & & \\
Downvotes : Editors' Picks Pinned & -0.0095015 & 0.0040785 & -2.33 & $0.019824$* \\
Rev. Downvotes : Editors' Picks Pinned & -0.0041419 & 0.0040755 & -1.02 & $0.30949$. \\
Upvotes : Editors' Picks Pinned & -0.0129699 & 0.0040824 & -3.18 & $0.001488$** \\
Relative Votes : Editors' Picks Pinned & -0.0118289 & 0.0040789 & -2.90 & $0.003731$** \\
Rev. Chronological : Editors' Picks Pinned & -0.0011811 & 0.0040763 & -0.29 & $0.77200$. \\
Chronological : Editors' Picks Pinned & -0.0030279 & 0.0040753 & -0.74 & $0.45748$. \\
Pred. Editors' Picks (LR) : Editors' Picks Pinned & -0.0146133 & 0.0040888 & -3.57 & $0.000352$*** \\
Pred. Editors' Picks (XGB) : Editors' Picks Pinned & -0.0147416 & 0.0041030 & -3.59 & $0.000327$*** \\
Pred. Upvotes (NBR) : Editors' Picks Pinned & -0.0153412 & 0.0040968 & -3.74 & $0.000181$*** \\
Pred. Upvotes (XGB) : Editors' Picks Pinned & -0.0130962 & 0.0040920 & -3.20 & $0.001372$** \\
\end{tabular}}
\label{tab:results-smog_N}
\end{table}

\section{Pinned Comment Time Distribution}
\label{timedist}

\begin{figure}[h]
    \centering
    \includegraphics[]{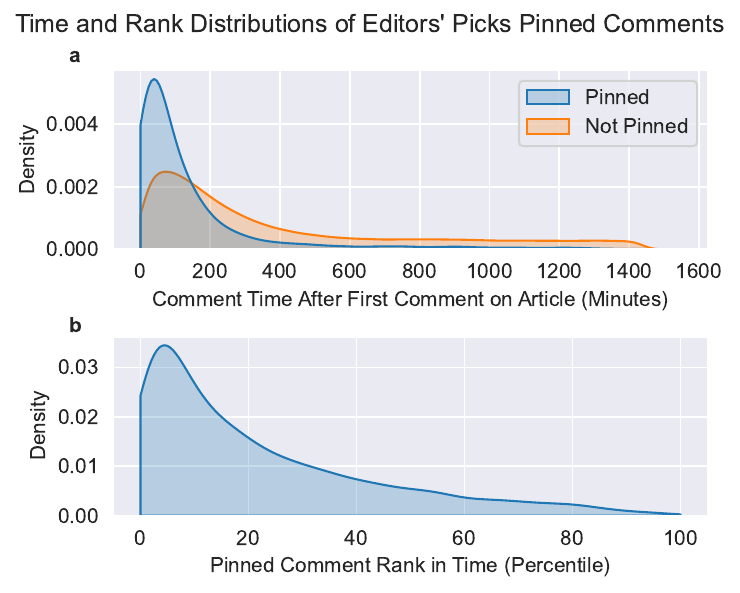}
    \caption{Kernel density estimate plots for the distributions of: a) Comment time after first comment on article for Editors' Picks and regular comments. b) The rank (as a percentile) of how early in time within an article discussion section that pinned comments are posted.}
    \label{fig:timedist}
\end{figure}

Here we show in Figure \ref{fig:timedist} how comments selected as Editors' Picks and pinned to the top of discussions are typically posted early in the discussion, both in absolute terms, and relative to other comments in the article comment section. Median time after first comment for Editors' Picks = $66$ minutes, other comments = $234$ minutes. Median time ranking percentile for Editors' Picks = $14\%$, other comments = $50\%$.

\end{document}